\pdfoutput=1
\documentclass[conference,compsoc]{IEEEtran}
\IEEEoverridecommandlockouts
\usepackage[T1]{fontenc}
\usepackage[utf8]{inputenc}
\usepackage{amsmath,amssymb}
\usepackage{booktabs}
\usepackage[font=footnotesize]{caption}
\usepackage{graphicx}
\usepackage{tikz}
\usepackage{tcolorbox}
\tcbuselibrary{breakable}
\usetikzlibrary{positioning,fit,arrows.meta,calc,backgrounds}
\usepackage{xcolor}
\usepackage{array}
\usepackage{multirow}
\usepackage{microtype}
\usepackage{enumitem}
\usepackage{xspace}
\usepackage{algorithm}
\usepackage{algpseudocode}
\usepackage{url}
\usepackage{cite}
\usepackage[hidelinks,hypertexnames=false,pdftitle={MirrorGuard: Defending LiDAR Perception against Passive Mirror Attacks},pdfauthor={Aymen Bouferroum, Ildi Alla, Selma Yahia, Vincent Lenders, Valeria Loscri},pdfcreator={},pdfproducer={}]{hyperref}
\usepackage[capitalise]{cleveref}
\crefname{figure}{Fig.}{Figs.}
\Crefname{figure}{Fig.}{Figs.}
\crefname{table}{Table}{Tables}
\Crefname{table}{Table}{Tables}
\crefname{section}{Section}{Sections}
\Crefname{section}{Section}{Sections}
\crefname{equation}{Eq.}{Eqs.}
\Crefname{equation}{Eq.}{Eqs.}
\crefname{algorithm}{Algorithm}{Algorithms}
\crefname{appendix}{Appendix}{Appendices}
\Crefname{appendix}{Appendix}{Appendices}

\usepackage{etoolbox}
\usepackage{amsmath,amssymb}
\usetikzlibrary{arrows.meta,calc,positioning}

\usetikzlibrary{arrows.meta,calc,positioning,decorations.markings}
\definecolor{cPath}{RGB}{0,114,178}
\definecolor{cPhantom}{RGB}{213,94,0}
\definecolor{cCar}{RGB}{120,120,120}
\definecolor{cObj}{RGB}{200,120,30}
\tikzset{
  midarrow/.style={decoration={markings, mark=at position #1 with {\arrow{Stealth[length=5pt]}}}, postaction={decorate}},
  mgpath/.style  ={cPath,   line width=0.7pt},
  mgph/.style    ={cPhantom,line width=0.8pt, dashed, -{Stealth[length=5pt]}},
  mgmirror/.style={black, line width=2.0pt},
  mgpt/.style    ={circle, fill=black, inner sep=1.2pt},
  mglbl/.style   ={font=\footnotesize},
  mgcar/.style   ={draw=cCar, fill=cCar!40, rounded corners=1pt, minimum width=10mm, minimum height=5mm, inner sep=0pt},
  mgobj/.style   ={draw=cObj, fill=cObj!45, rounded corners=1pt, minimum width=10mm, minimum height=5mm, inner sep=0pt},
  mgghost/.style ={draw=cPhantom, dashed, fill=cPhantom!15, rounded corners=1pt, minimum width=10mm, minimum height=5mm, inner sep=0pt},
}

\newcommand{\system}{MirrorGuard}
\newcommand{\carla}{CARLA}
\newcommand{\tof}{time-of-flight}
\providecommand{\degree}{\ensuremath{^{\circ}}}

\newcommand{\tabfont}{\scriptsize\setlength{\tabcolsep}{3pt}}

\definecolor{mgblue}{HTML}{002E99}\definecolor{mgred}{HTML}{D32F2F}\definecolor{mgtan}{HTML}{F5BC7A}
\definecolor{mglightblue}{HTML}{7E8EC1}\definecolor{mgsalmon}{HTML}{F2AD94}\definecolor{mgbrown}{HTML}{D97E3D}

\newcommand{\circledf}[1]{\tikz[baseline=(char.base)]\node[shape=circle,fill=black,inner sep=0.7pt,text=white] (char) {#1};}

\begin{document}

\title{\system{}: Defending LiDAR Perception against Passive Mirror Attacks}

\author{\IEEEauthorblockN{Aymen Bouferroum\textsuperscript{1}, Ildi Alla\textsuperscript{2}, Selma Yahia\textsuperscript{2}, Vincent Lenders\textsuperscript{2}, Valeria Loscri\textsuperscript{1}}
\IEEEauthorblockA{\textsuperscript{1}Inria Lille-Nord Europe, Lille, France\\
\textsuperscript{2}University of Luxembourg, Luxembourg}}

\maketitle
\bstctlcite{IEEEexample:BSTcontrol}

\begin{abstract}

Mirror-based LiDAR attacks passively redirect automotive LiDAR's beams to create phantom objects in the vehicle perception. Recent work has demonstrated that these attacks may  cause vehicles to brake for a phantom where nothing stands. Every phantom return is a genuine \tof{} measurement, so current spoofing defenses are not able to differentiate phantoms from real objects. %
We present \system{}, a physics-informed defense that identifies and removes mirror-induced phantom returns before they affect downstream perception.
\system{} tests three consequences of the reflected path. Every phantom bearing crosses the mirror's finite aperture, its opaque surface blocks any direct return beyond it, and the two reflections off the mirror dim the phantom. \system{} identifies a phantom by these cues over multiple scans and removes the reflected beams before it reaches any detector or occupancy grid.
On
real-world recordings of this attack, \system{} removes 96.31\% of the phantom's returns at 98.30\% precision, with 99.99\% of non-phantom returns untouched.
In a closed loop with the \carla{} driving simulator and the open-source stack (Autoware), a vehicle without the defense halts for an obstacle that is not present and a following car crashes into its rear. With \system{}, no vehicle
produces a halt or a crash.
\system{} runs at 59\,ms mean per scan on a CPU and is a software-only solution that does not require new hardware or environmental training. %

\end{abstract}

\section{Introduction}
\label{sec:intro}

\suppressfloats[t] %
\begin{figure}[t]
\centering
\includegraphics[width=\columnwidth]{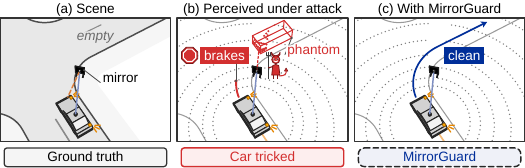}
\caption{A vehicle turning right past a roadside mirror. (a)~The reference attack, in which the mirror reflects the LiDAR's beams back to the vehicle. (b)~The vehicle braking for its own phantom.
(c)~\system{}
removes the phantom returns so the vehicle turns.}
\label{fig:overview}
\end{figure}

An autonomous vehicle's driving decisions rest on what its sensors report. Light Detection and Ranging (LiDAR) measures the \tof{} of its own laser pulses and builds the three-dimensional point cloud that feeds obstacle detection, localization and planning~\cite{leong2024lidar}.
Driving stacks deployed on real vehicles~\cite{kato2018autoware} treat these points as \textit{ground truth} and \textit{brake} for any obstacle they place in the planned path. An adversary who corrupts the point cloud therefore steers the vehicle without touching it.

Over the past decade, LiDAR spoofing research has focused on active attacks, in which external laser emitters inject forged optical pulses to create or erase obstacles~\cite{petit2015remote,shin2017illusion,cao2019adversarial,sato2025realism}. These attacks introduce foreign optical energy, and synchronized injection \textit{requires precise signal timing}. Existing countermeasures detect the traces these injections leave, such as violations of free-space and occlusion constraints~\cite{sun2020towards}, missing object shadows~\cite{hau2021shadowcatcher} and temporal inconsistency~\cite{cho2023adopt}. Newer sensors randomize pulse timing and fingerprint their pulses, which makes synchronized injection impractical~\cite{sato2024newgen}.

Recent work by Yahia et al.~\cite{yahia2025seeing} showed that LiDAR perception can be manipulated without injecting any signal. An adversary places planar mirrors at the roadside and redirects the sensor's own laser beams toward a real object (\cref{fig:overview}a). The LiDAR receives these reflected pulses and reads their total \tof{} as a direct-path distance, so a copy of the object, the \textit{phantom}, appears where nothing stands (\cref{fig:overview}b). The attack costs tens of dollars in materials. In real-vehicle experiments, the phantom propagated through the perception and planning pipeline and caused unnecessary braking~\cite{yahia2025seeing}. Every phantom return is a genuine \tof{} measurement of the sensor's own pulse~\cite{yahia2025seeing}, so neither the injection traces nor the pulse fingerprints reveal it. We call this attack the \textit{reference attack}.

Defending against this attack therefore requires distinguishing direct-path returns from those produced by reflections. A thermal camera can identify the phantom by its missing heat signature, but needs \textit{extra hardware} and is weakened by ambient heat, rain or dirt on the lens, and small or distant mirrors~\cite{yahia2025seeing}. The intensity detector~\cite{park2024intensity} flags detected objects by their share of bright returns and misses the dim mirror phantom~\cite{yahia2025seeing}. A per-point darkness test catches the phantom only by also flagging dark asphalt, shadows and low-albedo surfaces~\cite{yahia2025seeing}. Among reflection-aware methods, ReAL~\cite{solanki2025real} detects reflective attacks on LiDAR range and width measurements with a learned classifier. It was tested with a single-line 2D LiDAR at short range in the lab and \textit{does not identify} which returns are reflected. Full-waveform~\cite{ikeda2026ghostfwl}, multibounce~\cite{henley2023multibounce} and multi-echo~\cite{koch2017detection} methods need data that standard single-return LiDARs do not produce. Yang and Wang~\cite{yang2011solving} filter specular reflections from repeated scans of a static scene, whereas an autonomous vehicle has a limited window to spot a roadside mirror. \Cref{tab:related} compares these defenses.

\DeclareRobustCommand{\bgfull}{\tikz[baseline=-0.6ex]\fill (0,0) circle (2.1pt);}
\DeclareRobustCommand{\bgpart}{\tikz[baseline=-0.6ex]{\draw[line width=0.5pt] (0,0) circle (2.1pt);\fill (0,0) -- (90:2.1pt) arc (90:270:2.1pt) -- cycle;}}
\DeclareRobustCommand{\bgnone}{\tikz[baseline=-0.6ex]\draw[line width=0.5pt] (0,0) circle (2.1pt);}
\begin{table}[t]
\caption{Defenses against LiDAR point-cloud deception and specular artifacts, compared on the properties a fleet deployment needs. \bgfull{} holds, \bgpart{} partial, \bgnone{} absent or not reported, $^\dagger$ closed loop in simulation.}
\label{tab:related}
\centering
\tabfont
\begin{tabular}{@{}lcccccccccc@{}}
\toprule
 & \multicolumn{2}{c}{Scope} & \multicolumn{4}{c}{Deployment} & \multicolumn{4}{c}{Evaluation} \\
\cmidrule(lr){2-3}\cmidrule(lr){4-7}\cmidrule(lr){8-11}
 & \rotatebox{90}{Mirror phantoms} & \rotatebox{90}{Per-point} & \rotatebox{90}{Stock LiDAR only} & \rotatebox{90}{CPU real-time} & \rotatebox{90}{No scene prior} & \rotatebox{90}{Physics-based} & \rotatebox{90}{\shortstack[l]{Physical\\vehicle attack}} & \rotatebox{90}{Closed loop} & \rotatebox{90}{White-box attacker} & \rotatebox{90}{\shortstack[l]{Transfers without\\retraining}} \\
\midrule
\addlinespace[2pt]
\multicolumn{11}{@{}l}{\textit{Reflection filters}} \\
Yang and Wang~\cite{yang2011solving} & \bgfull & \bgfull & \bgfull & \bgnone & \bgpart & \bgfull & \bgnone & \bgpart & \bgnone & \bgnone \\
Koch et al.~\cite{koch2017detection} & \bgfull & \bgfull & \bgpart & \bgnone & \bgfull & \bgfull & \bgnone & \bgnone & \bgnone & \bgnone \\
Henley et al.~\cite{henley2023multibounce} & \bgfull & \bgfull & \bgnone & \bgnone & \bgfull & \bgfull & \bgnone & \bgnone & \bgnone & \bgnone \\
Ghost-FWL~\cite{ikeda2026ghostfwl} & \bgfull & \bgfull & \bgnone & \bgnone & \bgfull & \bgnone & \bgnone & \bgpart & \bgnone & \bgnone \\
\addlinespace[2pt]
\multicolumn{11}{@{}l}{\textit{Reflective-attack defenses}} \\
Thermal camera~\cite{yahia2025seeing} & \bgfull & \bgnone & \bgnone & \bgnone & \bgfull & \bgfull & \bgfull & \bgnone & \bgnone & \bgnone \\
ReAL~\cite{solanki2025real} & \bgpart & \bgnone & \bgfull & \bgpart & \bgpart & \bgnone & \bgpart & \bgnone & \bgnone & \bgnone \\
\midrule
\system{} (this work) & \bgfull & \bgfull & \bgfull & \bgfull & \bgfull & \bgfull & \bgfull & \bgfull\rlap{$^\dagger$} & \bgfull & \bgfull \\
\bottomrule
\end{tabular}
\end{table}

The mirror attack raises \textit{two challenges}. First, the phantom returns must be identified without prior knowledge of the mirror's location, orientation or the objects it reflects. Second, recognizing the phantom is not enough, because its returns populate the occupancy grid and can stop the vehicle even when the detector labels them only as unknown~\cite{yahia2025seeing}. The phantom returns themselves \textit{must therefore be removed from the point cloud}, without losing legitimate returns and at the pace of the sensor. A defense placed on the returns, before any detector, protects whatever detector the stack runs. Consuming only the data the sensor ships, it can reach vehicles already on the road without adding a sensor.

To address these challenges, we present \system{}, a software-only defense built on three consequences of reflection. Every reflected beam must intersect the mirror's finite \emph{aperture}, and because the mirror is \emph{opaque}, a return seen through it cannot be a direct return. The phantom is \emph{dark} against the surfaces around it at the same range, and the defense uses this darkness to confirm a phantom only on structures seen through a suspected mirror. \system{} finds suspected mirrors in each scan as large flat surfaces and combines the three consequences into a filter on the sensor's point stream that \textit{identifies} and \textit{removes} the phantom returns (\cref{fig:overview}c). Only the returns in the directions a suspected mirror covers wait for confirmation over a few scans, and every other return passes as soon as its scan is processed.

We evaluate \system{} on real recordings and in a reflection-exact simulation in \carla{}~\cite{dosovitskiy2017carla}, an open-source driving simulator, where the open-source stack Autoware~\cite{kato2018autoware} drives the vehicle. On the nine real-world (RW) recordings of the \textit{reference attack}~\cite{yahia2025seeing}, \system{} removes 96.31\% of the phantom vehicle's returns at 98.30\% precision and leaves 99.99\% of the 22.9 million non-phantom returns untouched. In every closed-loop attack drive without the defense, the planner stops for an obstacle that \textit{does not exist}, and when a car follows, it \textit{crashes} into the vehicle's rear. With \system{} no reference-attack drive produces a halt or crash. \system{} runs at 59\,ms mean per scan on a CPU.
Five published detectors trained on the labeled RW recordings take 3.32 to 11.23 times as long per scan on the CPU, and none reaches both \system{}'s recall and its precision on average over seeds.
Trained on simulation, they remove far more non-phantom than phantom returns on the RW recordings, while \system{} trains no model and runs with the same design values in simulation and on the real sensor.

In summary, this work makes the following contributions:
\begin{itemize}[leftmargin=1.2em,itemsep=1pt,topsep=2pt]
    \item We formalize phantom removal per return and derive three \textit{physical constraints} that every mirror-induced return obeys at any planar board and for any reflected object. The return's bearing crosses the board's finite aperture, the opaque board rules out a direct return beyond it on that bearing, and the double pass attenuates the return's power by the board's reflectance squared.

    \item We design \system{}, to our knowledge the \textit{first software-only, point-level} defense that identifies and removes the phantom returns of a physical passive-mirror attack while nearly every real return stays untouched. On the stock single-return stream it removes the confirmed returns \textit{before any consumer}, so one filtered cloud serves every detector and the occupancy grid.
    \item Nothing in \system{} is learned. Every value is set once and never adjusted per recording, site, environment or attack, and the values tied to the sensor come from the sensor itself without labels, so a new deployment needs \textit{no hand labels and no new real recordings of the attack}. We state the attack set \system{} covers and analyze an attacker who knows every parameter.

\end{itemize}

\section{Background and Problem Definition}
\label{sec:background-lidar}
\subsection{%
LiDAR and Mirror-Induced Reflections} Automotive spinning LiDAR sensors estimate the range of surrounding surfaces from the round-trip time of emitted
laser pulses. A scan is stored as a \textit{destaggered} beam-by-azimuth grid with $H$ beams and $W$ columns. The beams of each column fire at slightly different azimuths, and the destaggered image shifts each beam's samples to a common azimuth grid. Taking the sensor origin $O$ as the coordinate origin, the point cloud acquired at time $t$ is
\begin{equation}
    P_t = \{p_c=r_cb_c \mid c\in\mathcal{V}_t\},
    \label{eq:scan}
\end{equation}
where $\mathcal{V}_t\subseteq[H]\times[W]$ is the
set of valid return cells, $b_c$ is the unit beam direction, and $r_c$ is the measured range. Each return carries a calibrated reflectivity $I_c$, scaled so that $I_c{=}1$ is a 100\,\% Lambertian target~\cite{ouster2026reflectivity}. In the single-return profile considered in this work, the sensor provides one echo per pulse, without additional returns or waveform information.

A planar mirror can redirect an emitted pulse
toward a physical surface, so the sensor reconstructs the return at a virtual location.
Let $x_m$ denote the point where the emitted beam $b_c$ intersects the mirror, and let $q$ be the physical point reached after reflection. The optical path is reciprocal, and the measured range and reconstructed position are
\begin{equation}
    r_m=\|x_m\|+\|q-x_m\|,
    \qquad p=r_m b_c.
    \label{eq:phantom}
\end{equation}
For an ideal planar mirror through $m_0$ with unit normal $\hat n$, the reconstructed point coincides with the virtual image of $q$,\looseness=-1
\begin{equation}
    p=q'=q-2\bigl((q-m_0)^\top\hat n\bigr)\hat n.
    \label{eq:image}
\end{equation}
Reflections from an extended physical object therefore produce a spatially coherent phantom behind the mirror plane, even though no corresponding object occupies that location. These phantom returns originate from legitimate LiDAR pulses and enter the perception pipeline as ordinary
returns. They can generate \textit{false} object detections or \textit{spurious} occupancy-grid cells that influence vehicle planning.

\subsection{Threat Model}
\label{sec:threat}
We consider a physical-world adversary seeking to induce
unnecessary braking or stopping by placing one static planar mirror board at the roadside, which redirects the sensor's beams and creates a phantom obstacle on an autonomous vehicle's intended driving path. The attack is passive, with no emissions, no electronics, no synchronization with the sensor and only roadside access, leaving the vehicle's LiDAR, firmware, perception and planning uncompromised, with no access to internal states or sensor measurements. The adversary controls the board's placement, orientation, dimensions and surface properties, and may choose the reflected object.

\begin{figure}[!t]
\centering
\includegraphics[width=\columnwidth]{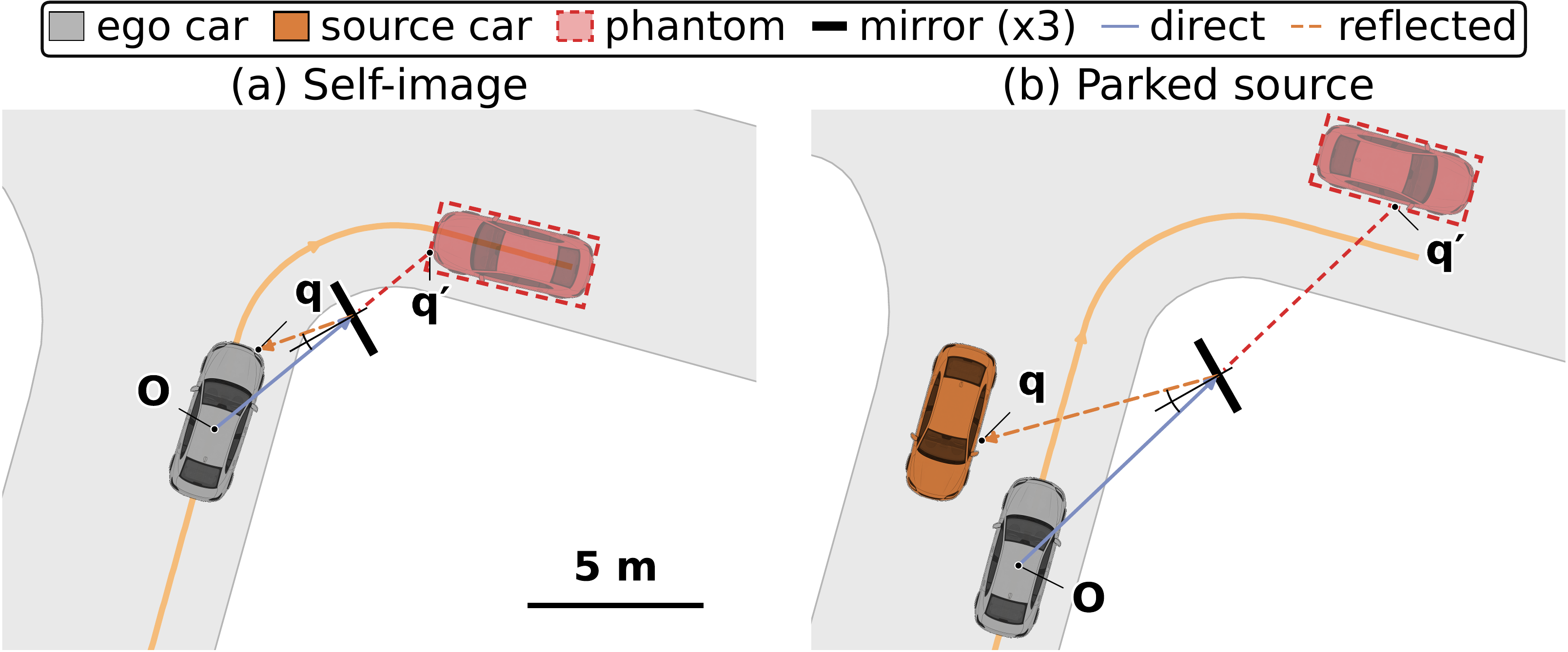}
\caption{The two attack scenarios in top view, drawn from one simulated scan each. $O$~sensor, $q$~physical source, $q'$~phantom position. The dashed line is the continuation to $q'$. The mirror bar is enlarged for visibility.}

\label{fig:threat}
\end{figure}

We consider two scenarios (Fig.~\ref{fig:threat}). In the \emph{self-image} scenario (Fig.~\ref{fig:threat}a), the board reflects the approaching vehicle itself, as demonstrated by Yahia et al.~\cite{yahia2025seeing}. In the \emph{parked-source} scenario (Fig.~\ref{fig:threat}b), the board reflects a parked vehicle, so the adversary chooses the source and with it the phantom's geometry. In both scenarios, the adversary arranges the physical scene before the vehicle approaches but has no control over its trajectory, speed or LiDAR acquisition. We consider both conventional attacks, in which the board configuration is selected independently of the defense, and adaptive attacks, in which the adversary knows the complete design and every parameter value of \system{} and chooses configurations intended to evade phantom removal.
In scope is phantom addition in a single drive-by,
and out of scope are object hiding, actuated
or convex mirrors and weather or lighting effects.

\subsection{Problem Formulation}
\label{sec:problem}

The virtual-image formation of Eq.~\eqref{eq:image} introduces phantom returns into the point cloud. For a board configuration $m\in\mathcal{A}$, where $\mathcal{A}$ denotes the set of physically admissible configurations, the resulting scan admits the partition
\begin{equation}
\begin{aligned}
    P_t &= R_t(m)\,\dot{\cup}\,G_t(m)
           \,\dot{\cup}\,\Phi_t(m),\\
    \Phi_t(m) &=
    \Phi_t^{\mathrm{obj}}(m)\,\dot{\cup}\,
    \Phi_t^{\mathrm{bg}}(m).
\end{aligned}
\label{eq:partition}
\end{equation}
Here, $R_t(m)$ comprises direct-path returns,
$G_t(m)$ is the \textit{glint} from the board surface, and $\Phi_t(m)$ collects returns reflected at the board, split into \textit{phantom-object} returns $\Phi_t^{\mathrm{obj}}(m)$ that form the perceived obstacle and \textit{phantom-background} returns $\Phi_t^{\mathrm{bg}}(m)$ from background structure seen through the board. Although these sets are distinguished by their physical origins, the defender observes only their union $P_t$, without knowledge of the underlying partition. Within $\mathcal{A}$, four conditions bound a covered subset $\mathcal{A}_{\mathrm{cov}}\subseteq\mathcal{A}$ that a solution must remove, namely \circledf{1}~a board large enough to appear planar in range, \circledf{2}~bright against its surroundings, \circledf{3}~the phantom within a bounded depth behind it, and \circledf{4}~the phantom in view over consecutive scans.\looseness=-1

The problem is to identify the returns forming phantom obstacles while preserving legitimate scene information. We express this as a point-level decision function $d_t:P_t\rightarrow\{0,1\}$, where $d_t(p)=1$ indicates that return $p$ should be removed. The removal set and filtered point cloud are
\begin{equation}
\begin{aligned}
    \widehat{\Phi}_t
    &=\{p\in P_t\mid d_t(p)=1\},\\
    \widetilde{P}_t
    &=P_t\setminus\widehat{\Phi}_t.
\end{aligned}
\label{eq:filter}
\end{equation}
Processing is causal at output time with a bounded release hold~$\tau_{\mathrm{hold}}$. Held returns pass within the hold plus two more scans and their processing, and all other returns pass once their scan is processed. A correct solution removes all phantom-object but no direct-path or glint returns, so the ideal removal set satisfies
\begin{equation}
    \Phi_t^{\mathrm{obj}}(m)
    \subseteq \widehat{\Phi}_t
    \subseteq \Phi_t(m).
\label{eq:ideal}
\end{equation}
Object recall measures the first inclusion and whole-frame precision the second.
Reflected
background returns may also be removed without violating this condition.

The challenge is that the partition in Eq.~\eqref{eq:partition} is not directly observable. Both direct-path and reflected returns originate from genuine LiDAR pulses, and their reconstructed coordinates do not reveal the propagation paths.
To detect phantoms, a solution may read only the sensor's own scans, that is, range, bearing and calibrated reflectivity, and to release held returns it may also use the vehicle pose the platform's localization publishes, without a camera, radar, map or learned model. The physical relationships that distinguish reflected returns from direct observations are examined in \cref{sec:invariants}.\looseness=-1

\section{Physical Constraints %
}
\label{sec:invariants}
\subsection{Finite-Aperture Geometry}
\label{sec:physics:aperture}

The virtual-image geometry
imposes a spatial constraint on mirror-induced measurements. Since reflection occurs at a finite surface, the bearing of a phantom return must intersect the reflective aperture before reaching its reconstructed position. This restricts phantom measurements to the region projected through the mirror's boundaries, regardless of the location or geometry of the reflected object.

We model the reflective surface of mirror $m$ as a rectangular aperture $\mathcal{B}_m$, with center $m_0$, unit normal $\hat n$, orthonormal in-plane directions $\hat u$ and $\hat v$, and half-widths $h_u$ and $h_v$:\looseness=-1
\begin{equation}
\begin{aligned}
\mathcal{B}_m = \bigl\{x\in\mathbb{R}^3 \mid\;&
(x-m_0)\cdot\hat n=0,\\
&|(x-m_0)\cdot\hat u|\leq h_u,\\
&|(x-m_0)\cdot\hat v|\leq h_v
\bigr\}.
\end{aligned}
\label{eq:aperture}
\end{equation}
For each observed return $p\in P_t$, the corresponding LiDAR bearing intersects the mirror plane at $x_m(p)=\lambda_m(p)p$, where the intersection parameter is given by
\begin{equation}
\lambda_m(p)=
\frac{m_0\cdot\hat n}{p\cdot\hat n},
\qquad
x_m(p)=\lambda_m(p)p,
\label{eq:plane_intersection}
\end{equation}
for $p\cdot\hat n\neq0$. A positive $\lambda_m(p)$ places the intersection along the forward bearing, while $0<\lambda_m(p)<1$ indicates that the mirror plane lies between the sensor and the reconstructed return. A return is consistent with reflection through $m$ only if the intersection lies within the aperture and precedes the apparent return. We express these conditions through
\begin{equation}
\begin{aligned}
\chi_m(p) \equiv\;&
\bigl|(x_m(p)-m_0)\cdot\hat u\bigr|\leq h_u\\
&{}\land
\bigl|(x_m(p)-m_0)\cdot\hat v\bigr|\leq h_v,\\
\zeta_m(p) \equiv\;&
0<\lambda_m(p)<1,\\
\eta_m(p) \equiv\;&
\chi_m(p)\land\zeta_m(p).
\end{aligned}
\label{eq:aperture_condition}
\end{equation}
The predicate $\chi_m$ enforces intersection with the finite mirror surface, while $\zeta_m$ requires the reconstructed return to lie beyond the mirror plane along its bearing. Together, they define the geometric support of phantom measurements. Under the ideal reflection model,
\begin{equation}
p\in\Phi_t(m)
\;\Longrightarrow\;
\eta_m(p)=1.
\label{eq:aperture_necessary}
\end{equation}
\subsection{Occlusion and Path Constraints}
\label{sec:physics:opacity}
The opacity of the mirror provides a constraint on the origin of LiDAR returns. A beam intersecting the reflective aperture cannot reach surfaces located behind the mirror. Therefore, a return reconstructed beyond the mirror plane along the same bearing is inconsistent with direct-path propagation and instead indicates a reflected optical path.

\begin{figure*}[t]
\centering
\input{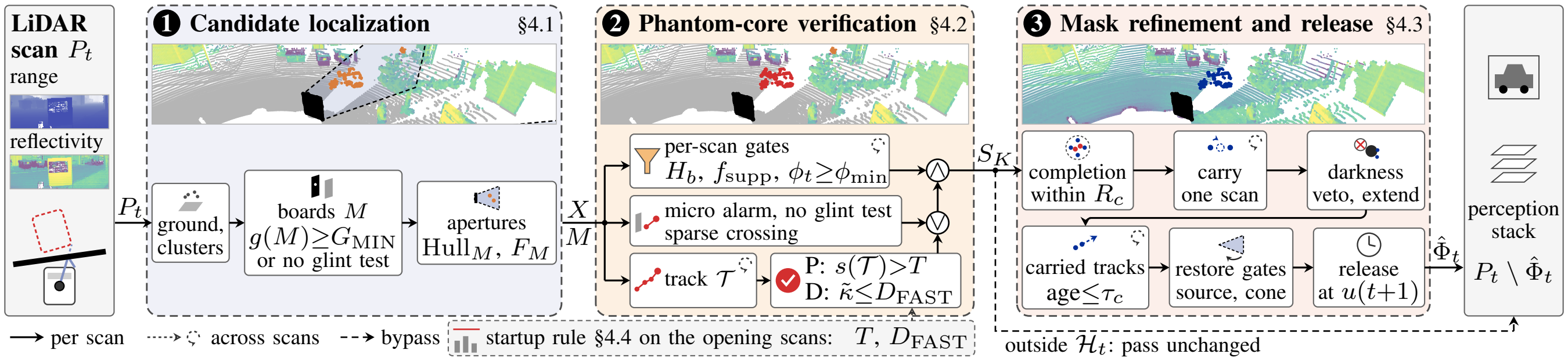}\unskip
\caption{Architecture of \system{} on one scan. Phase I
locates boards and structures seen through them (orange), Phase II
confirms a phantom core across scans (red), and Phase III
grows the removal mask (blue), passing the rest to the perception stack.}
\label{fig:pipeline}
\end{figure*}

Consider a return $p\in P_t$ satisfying $\eta_m(p)=1$, with the corresponding intersection $x_m(p)\in\mathcal{B}_m$. Since the mirror lies between the sensor and the reconstructed point, the direct path to $p$ intersects the opaque surface. Under the ideal mirror model, this excludes a direct-path interpretation, so $\eta_m(p)=1\Longrightarrow p\notin R_t(m)$.
Combining this constraint with the partition
in Eq.~\eqref{eq:partition} yields
\begin{equation}
\begin{aligned}
\mathcal{E}_t(m)
&=\{p\in P_t\mid\eta_m(p)=1\},\\
\mathcal{E}_t(m)
&\subseteq\Phi_t(m).
\end{aligned}
\label{eq:opacity_support}
\end{equation}

This exclusion depends on the physical mirror boundary being known accurately. In practice, uncertainty in surface localization, finite beam divergence, and partial intersections near mirror edges can make the distinction ambiguous.

\subsection{Radiometric Effects of Double-Pass Reflection}
\label{sec:physics:radiometry}

Mirror-induced returns exhibit a radiometric signature arising from the two interactions of the LiDAR pulse with the reflective surface. The emitted pulse is first redirected toward the physical object, and the backscattered signal undergoes a second reflection before reaching the sensor. Consequently, both the outgoing and returning optical paths are affected by the mirror's reflectance. Let $\rho_m(\theta)\in[0,1]$ denote the effective power reflectance of mirror $m$ at incidence angle $\theta$. Under reciprocal propagation, the two interactions introduce a multiplicative power-transfer factor
\begin{equation}
    T_m =
    \rho_m(\theta_{\mathrm{out}})
    \rho_m(\theta_{\mathrm{ret}}).
    \label{eq:mirror_transfer}
\end{equation}
For coincident outgoing and returning paths, the two angles are equal, $\theta_{\mathrm{out}}=\theta_{\mathrm{ret}}=\theta_m$, and hence $T_m=\rho_m^2(\theta_m)$.

Let $P_{\mathrm{rx}}^{(0)}(q)$ denote the power that would be received from the same surface $q$ under equivalent propagation geometry without losses at the mirror. The received power of the corresponding phantom return is then
\begin{equation}
    P_{\mathrm{rx}}^{\mathrm{ph}}(q,m)
    = T_m\,P_{\mathrm{rx}}^{(0)}(q)
    = \rho_m^2(\theta_m)
      P_{\mathrm{rx}}^{(0)}(q).
    \label{eq:phantom_power}
\end{equation}
This relation isolates the attenuation introduced by the mirror from the radiometric properties of the reflected surface.

For calibrated LiDAR reflectivity, the corresponding relationship can be approximated as
\begin{equation}
    I_{\mathrm{ph}}(q,m)
    \approx
    \rho_m^2(\theta_m)
    \varepsilon(q,m)\,I_q,
    \label{eq:phantom_reflectivity}
\end{equation}
where $I_q$ represents the reference reflectivity of the physical surface and $\varepsilon(q,m)$ captures residual effects of incidence geometry, propagation conditions, and sensor calibration. The mirror only attenuates the return, so the phantom's reflectivity is set by the imaged surface and reduced by the double pass.

The geometric and radiometric properties established in this section provide complementary evidence for identifying mirror-induced phantom returns. Finite-aperture geometry constrains their spatial distribution, mirror opacity restricts their possible propagation paths, and double-pass reflection introduces characteristic radiometric attenuation.

\section{\system{} Design}
\label{sec:design}

Building on the physical properties of Section~\ref{sec:invariants}, we introduce \system{}, a physics-informed defense that identifies and removes mirror-induced phantom returns before they affect downstream perception, covering the attack set $\mathcal{A}_{\mathrm{cov}}$ of \cref{sec:problem}.
\system{} operates as a per-sensor LiDAR filter, processing each sensor's point cloud before multi-sensor aggregation. As shown in \cref{fig:pipeline},
the defense follows three stages. First, it localizes candidate boards and identifies measurements geometrically consistent with reflected propagation. Second, it verifies these candidates using opacity, radiometric, and temporal evidence. Finally, it removes confirmed phantom returns and forwards the filtered point cloud to the perception stack.

\subsection{Phase I: Candidate Localization}
\label{sec:design:board}

Phase~I identifies candidate mirror boards and LiDAR returns consistent with reflection through their apertures. We estimate the ground plane $\Pi_g$ using deterministic RANSAC~\cite{fischler1981random} and exclude its inliers. The remaining returns are clustered using DBSCAN~\cite{ester1996density}. For each cluster $X$, principal component analysis (PCA) of its covariance matrix estimates the surface normal $\hat n_X$ from the eigenvector associated with the smallest eigenvalue $\xi_{\min}(X)$.
Its square root $\sigma_X=\sqrt{\xi_{\min}(X)}$ is the cluster's thickness.

For a cluster $M$, we define the \emph{glint ratio} as
\begin{equation}
g(M)=
\frac{\operatorname{median}_{p\in M}I(p)}
{\operatorname{median}_{q\in N(M)}I(q)},
\label{eq:glint}
\end{equation}
where $N(M)$ holds the returns within the same range band as $M$, excluding $M$.
Geometrically admissible boards satisfy
\begin{equation}
\begin{aligned}
\Gamma(M)\equiv\;&
|M|\ge n_{M,\min}
\;\wedge\;\sigma_M<\sigma_{\max}\\
&{}\wedge\;|\hat n_{M,z}|<\nu_{\max}\\
&{}\wedge\;r_{M,\min}<\bar r_M<r_{M,\max},
\end{aligned}
\label{eq:board}
\end{equation}
where $\bar r_M$ is the horizontal distance to the cluster centroid. These constraints enforce minimum point support, planarity, near-vertical orientation, and an admissible distance from the sensor. A \textit{glint-accepted board} satisfies $\Gamma(M)$ and $g(M)\ge G_{\mathrm{MIN}}$ (the set $\mathcal{M}_t^G$), and a \textit{second family} satisfies $\Gamma(M)$ alone, without the glint test (the set $\mathcal{M}_t^V$). Glint-accepted boards $\mathcal M_t^G$ are used for opacity verification and temporal tracking, whereas $\mathcal M_t^V$ admits boards independently of their reflectivity to support the micro alarm and mask refinement.

For each candidate board $M$, the observed return directions estimate the reflective aperture $\mathcal B_m$ of \cref{sec:physics:aperture}
and define an angular hull $\mathrm{Hull}_M$, the convex hull of the return directions enlarged by $\mu$, and a pixel footprint $F_M$, the board's cells dilated by $k_F$ steps.
A cluster $X$ is \textit{seen through} $M$ when its centroid lies beyond the plane, its mean range exceeds the board's, and
\begin{equation}
\frac{|\{c\in X : c\in F_M\}|}{|X|}\ \ge\ \omega_{\min}.
\label{eq:overlap}
\end{equation}
Each see-through cluster behind a board becomes a candidate with its attributing board, footprint and plane.

The micro alarm admits smaller see-through clusters behind second-family boards at a lower persistence bound. A sparse rectangle detector fits rectangles to planar clusters and groups otherwise unclustered returns seen through them, feeding its own track score in Phase~II.

\subsection{Phase II: Phantom-Core Verification}
\label{sec:design:structure}
\label{sec:design:confirm}

Phase~II verifies candidate phantom structures using spatial, opacity, radiometric, and temporal evidence. Candidates sharing board cells form a structure $K$, with combined seed $S_K$, the union of their cells. Three per-scan gates assess each structure.
The \emph{band gate} requires at least a fraction $f_{\mathrm{band}}$ of seed returns with known height to lie within the height band $H_b$ above the ground plane.
The \emph{support gate} requires at least one associated board $M$ and its corresponding seed $S$, restricted to the footprint $F_M$, to satisfy
\begin{equation}
\frac{
|\{c\in S:h(c)\in H_b
\wedge c\in\mathrm{Region}(S)\}|
}{|S|}
\ge f_{\mathrm{supp}},
\label{eq:support}
\end{equation}
where $\mathrm{Region}(S)$ intersects $F_M$ with the row and column span of $S$.
The \emph{opacity gate} measures the fraction of seed returns whose apparent positions lie beyond
a glint-accepted board along an intersecting beam,
\begin{equation}
\begin{aligned}
\phi_t(S_K)=\frac{1}{|S_K|}
\bigl|\{c\in S_K:\;&
\exists M\in\mathcal M_t^G,\\
&b_c\in\mathrm{Hull}_M,\\
&0<\lambda_M(p_c)<1,\\
&d_M^{\mathrm{beam}}(p_c)\le D_{\mathrm{IMG}}
\}\bigr|,
\end{aligned}
\label{eq:opacity}
\end{equation}
where $\lambda_M$ is the plane-intersection parameter from Section~\ref{sec:physics:aperture} and $d_M^{\mathrm{beam}}(p_c) =r_c[1-\lambda_M(p_c)]$ is the beam depth
beyond the board plane. Opacity is accepted when $\phi_t(S_K)\ge\phi_{\min}$, through a track latch while the track of a previously linked glint-accepted board remains associated, or through a digest latch following an earlier successful opacity test for a board in the structure.

The source reflectivity $I_q$ in \cref{eq:phantom_reflectivity} is unknown to the filter, so radiometric evidence is evaluated relative to returns at comparable ranges. For a structure $K$ with mean range $\bar r_K$, let $B(K)$ contain non-ground returns within $\Delta r$ of $\bar r_K$, using the full non-ground population when its size does not exceed $n_{B,\min}$. The \emph{floored darkness statistic} is
\begin{equation}
\kappa_t(K)=
\frac{\operatorname{median}_{p\in K}I(p)}
{\max\!\left(
\operatorname{median}_{q\in B(K)}I(q),
I_{\mathrm{floor}}\right)},
\label{eq:kappat}
\end{equation}
where $I_{\mathrm{floor}}$ prevents normalization by an excessively small background reflectivity.
See-through clusters associated with glint-accepted boards are linked across scans by nearest-neighbor association in azimuth and range, without vehicle pose or per-scan gating. Each observation records the cluster size $n_c$, footprint overlap $\omega$, range $r$, and darkness statistic $\kappa$. For a track $\mathcal T$, let $n=|\mathcal T|$ and denote temporal medians by $\tilde{x}$. The track score is

\begin{equation}
s(\mathcal T)=
\frac{
\min(n,N_{\mathrm{cap}})
\,\tilde\omega\,\ln(1+\tilde n_c)
}{
\sqrt{\max(\tilde r,r_{\mathrm{floor}})}
\left[1+(\tilde\kappa/\kappa_0)^2\right]
}.
\label{eq:score}
\end{equation}
Two temporal confirmation paths are evaluated,
\begin{equation}
\begin{aligned}
\mathrm{P}:&\quad
n\ge N_P\;\wedge\;s(\mathcal T)>T,\\
\mathrm{D}:&\quad
n\ge N_{\mathrm{MIN}}
\;\wedge\;\tilde\kappa\le D_{\mathrm{FAST}}.
\end{aligned}
\label{eq:paths}
\end{equation}
Path~P requires more observations for brighter structures to exceed $T$, whereas Path~D accepts sufficiently persistent structures with low reflectivity ratios. The thresholds $T$ and $D_{\mathrm{FAST}}$ are established once per sensor by the label-free startup rule in \cref{sec:design:constants}.

The micro alarm maintains separate tracks of smaller see-through clusters associated with second-family boards. A track raises an alarm when its length, its duration and the temporal medians of reflectivity ratio, footprint overlap, range, distance from the associated plane, azimuth offset, range ratio, and height above the local lowest return satisfy fixed bounds (values in Appendix~\ref{app:constants}). The micro alarm also requires opacity verification.\looseness=-1

A fourth alarm, the sparse crossing, fires when the sparse detector's track score crosses its threshold. A structure is \textit{confirmed} at scan $t$ if its per-scan gates pass and one of its candidates $k$ has raised an alarm satisfying
\begin{equation}
a_k\le\tau_t+\tau_{\mathrm{hold}},
\label{eq:confirmation}
\end{equation}
where $a_k$ denotes the candidate's alarm time. Confirmed structures are forwarded to Phase~III for point-level filtering.

\subsection{Phase III: Mask Refinement and Release}
\label{sec:design:output}

Phase~III constructs the removal mask from the phantom structures confirmed in Phase~II, refines the selected returns using geometric and radiometric constraints,
and releases the filtered LiDAR scans.
For each confirmed candidate $k$, let $S_k$ denote its seed restricted to the associated board footprint $F_{M_k}$ from Phase~I. The \emph{core} expands this seed to recover adjacent phantom returns. The completion set is defined as
\begin{equation}
\begin{aligned}
\mathcal{W}_k=\operatorname{closure}\bigl(
S_k;\;&\mathcal N,\,
\mathrm{Box}_{w_c}(S_k),\\
&|r_{c'}-\tilde r_k|\le R_c,\\
&s_{M_k}(p_{c'})\ge b,\,
\mathrm{Adm}_{M_k}(c')
\bigr),
\end{aligned}
\label{eq:completion}
\end{equation}
where $\mathcal N$ is the range-image neighborhood of a cell, $\mathrm{Box}_{w_c}(S_k)$ is the row and column span of the seed enlarged by $w_c$ cells, and $\tilde r_k$ is its median range.
Additional returns must satisfy the range, signed plane-depth, and footprint-admission constraints associated with $M_k$, whereas the original seed enters the core without these tests. If all completion sets of a confirmed structure are empty, its entire seed $S_K$ is selected for removal.

To recover phantom returns during isolated detection gaps, a one-scan \emph{bracketed carry} $C_t$ is applied when no confirmed core exists at scan $t$, but cores are available at scans $t-2$, $t-1$, and $t+1$. Returns near the predicted position are selected within an adaptive radius $R_{\mathrm{carry}}=\operatorname{med}(d)+\alpha_c\hat\sigma(d)$, where $d$ contains prediction residuals from the last $W_{\mathrm{carry}}$ scan triples and $\hat\sigma$ is their normal-consistent median absolute deviation. The carry is suppressed when $R_{\mathrm{carry}}>r_{\mathrm{assoc}}$.

The \emph{darkness stage} $\mathcal D$ processes the combined core and carry returns. Components with insufficient support for reliable reflectivity estimation are retained. Components whose median reflectivity ratio reaches $\rho_{\mathrm{bright}}$ are discarded unless a short path through nearer returns connects them to a retained dark cell. Surviving components are expanded through $\mathcal N$ into neighbors whose reflectivity is at most $\rho_{\mathrm{dark}}$ times the background reflectivity in their range bins, whose height exceeds $h_{\mathrm{clip}}$, and whose range differs by at most $R_c$ from the preceding cell in the expansion.\looseness=-1

A rigid-template \emph{continuation tracker} follows the clusters of the refined removal set $B_t$. When no matching observation is found, returns near the motion-predicted template form a continuation set $E_t$. A track expires when the time since it last matched a cluster of $B_t$ exceeds $\tau_c$. The final removal mask is
\begin{equation}
\hat\Phi_t=\bigl(B_t\cup E_t\bigr)\setminus\bigl(V^{\mathrm{src}}_t\cup V^{\mathrm{cone}}_t\bigr),\quad B_t=\mathcal{D}(\mathrm{core}_t\cup C_t),
\label{eq:mask}
\end{equation}
where two restoration gates limit unintended removals. The source gate $V_t^{\mathrm{src}}$ restores continuation returns not observed through their associated board, while the cone gate $V_t^{\mathrm{cone}}$ restores returns outside the angular hull $\mathrm{Hull}_M$ of their attributed board. Returns without an observed board attribution are not subjected to these restoration tests.

The release mechanism accommodates temporal verification and bracketed carry. The core decision for scan $t$ becomes available at
$u(t)=\min\{u:\tau_u\ge\tau_t+\tau_{\mathrm{hold}}\}$, while its final removal mask is published at $u(t+1)$ to incorporate the subsequent scan required by bracketed carry.
Only returns within the \emph{held set} $\mathcal H_t$ are buffered. This set includes the footprints of glint-accepted boards observed during the preceding $\tau_{\mathrm{mem}}$, reprojected into scan $t$ using vehicle pose because the boards remain fixed in the world. It also includes the cells removed by every release decided within the same interval, and the entire set is widened by $k_{\mathrm w}$ cells. Returns within $\mathcal H_t$ are held until the removal mask for scan $t$ is published at $u(t+1)$ and are then released after applying $\hat\Phi_t$. All other returns are forwarded immediately.

\subsection{Setting the Parameters}
\label{sec:design:constants}
\system{} uses a set of constants that are fixed before deployment and never adjusted for individual recordings, environments, or attack configurations. We refer to these values as \emph{frozen}. No model training or labeled data are required. Apart from sensor-dependent quantities and the startup thresholds, all constants remain identical across scenes and datasets. The angular margin $\mu$ equals half the sensor's angular pitch, while $I_{\mathrm{floor}}$ is set to one Lambertian unit, the reference unit of calibrated LiDAR reflectivity corresponding to a Lambertian surface. Two constants follow explicit design rules. The glint threshold $G_{\mathrm{MIN}}$ exceeds the largest glint ratio observed in an attack-free simulated drive, while $D_{\mathrm{IMG}}$ exceeds the maximum simulated beam depth beyond the mirror board. The minimum confirmation length satisfies $N_{\mathrm{MIN}}=N_{\mathrm{cal}}$. The sparse-detector threshold follows a separate simulation-based rule, and the micro alarm bounds are listed in Appendix~\ref{app:constants}. The remaining constants are fixed design parameters, including the release hold $\tau_{\mathrm{hold}}$, which is specified in seconds independently of the sensor rate and $N_{\mathrm{MIN}}$.\looseness=-1

The confirmation thresholds $T$ and $D_{\mathrm{FAST}}$ are established through a label-free \emph{startup rule}, executed once per sensor. The first $N_w$ scans serve as a warm-up period, followed by a calibration window of $N_c$ scans. During calibration, candidate tracking proceeds without the glint gate or opacity test, allowing tracks to form behind every see-through candidate. Let $\mathcal E_{\mathrm{cal}}$ denote the eligible observations in this window. Each observation corresponds to a track at a particular scan, provided that the track has accumulated at least $N_{\mathrm{cal}}$ observations by that scan. The thresholds are computed as
\begin{equation}
T=\beta\max_{e\in\mathcal E_{\mathrm{cal}}}s(e),\qquad D_{\mathrm{FAST}}=\frac{1}{\beta}\min_{e\in\mathcal E_{\mathrm{cal}}}\tilde\kappa(e),
\label{eq:startup_calibration}
\end{equation}
where $s(e)$ and $\tilde\kappa(e)$ denote the track score and median darkness statistic at the corresponding scan, and $\beta$ is a fixed startup factor. If $\mathcal E_{\mathrm{cal}}=\varnothing$, the rule returns \emph{not-calibrated} and the filter does not start. Otherwise, both thresholds remain frozen during operation.
The sensitivity of \system{} to the design-rule and fixed design values is
detailed
in Appendix~\ref{app:results} (\cref{fig:sweep}). The experimental configurations and resulting parameter values are reported in the following section.

\section{Experimental Setup}
\label{sec:setup}

We evaluate \system{} in a \carla{}-based simulation~\cite{dosovitskiy2017carla} that reproduces mirror-induced reflections at the point level and on real-world LiDAR recordings of the reference attack~\cite{yahia2025seeing}. The simulation enables controlled variation of attack configurations and closed-loop driving, while the recordings provide physical attack measurements.

\subsection{Reflection-Exact Attack Simulation}
\label{sec:sim}

A point-level defense is judged return by return, so its evaluation needs the origin of every return, which the real recordings give only for the hand-labeled returns of a subset of scans. Our simulation gives this origin by construction, as the partition of \cref{eq:partition}. Beyond what a real test drive can cover safely or systematically, it extends the vehicle speed and the reflective area past the recorded values, varies the channel count, adds the parked-source scenario, and replays the drives with the reflection off as attack-free twins. It also covers an attacker who knows the defense and a closed loop in which a complete driving stack brakes for what it perceives and a following car can crash into the braking vehicle. For results to transfer to the real sensor, every phantom return must lie on the sensor's beam at the folded-path range (\cref{eq:phantom}) with the reflectivity of real phantoms, and the board must hide the returns behind it. \Cref{sec:eval:transfer} tests this transfer on the real recordings.

We take the three tilts ($30\degree{}$, $45\degree{}$ and $60\degree{}$) and the 0.60\,m$^2$ board of the reference attack~\cite{yahia2025seeing} and place the board at the distances where its model predicts a phantom. The reference attack's own \carla{} simulation adds a Gaussian cluster of points around a predicted phantom centroid to the native scan, with no beam structure, no reflectivity model and no removed return, which suits object-level effects on a driving stack but not a per-return defense. MirrorDrift~\cite{nagata2026mirrordrift}, an actuated-mirror attack on LiDAR SLAM, models the mirror with a virtual sensor at the mirror image of the real one (\cref{eq:image}). On a real LiDAR its simulated cloud lies at a Chamfer distance of 0.028\,m from the measured one, but it reflects recorded clouds outside a driving simulator and has no intensity model. We therefore bring this virtual sensor into \carla{} as an extension that provides what a per-return defense needs, beam placement, occlusion in both directions, origin labels and reflectivity, so that defenses against mirror attacks can be scored per return and tested in closed loop.

Because \carla{}'s ray casting assumes diffuse reflection and produces no mirror returns~\cite{yahia2025seeing}, the extension realizes the virtual sensor as a second LiDAR. A return of this virtual sensor is kept only if its bearing crosses the board aperture (\cref{sec:physics:aperture}), and it is placed on the vehicle's beam at the folded-path range. A nearer direct-path return on the same beam hides the phantom, and the opaque board hides the direct-path returns behind it (\cref{sec:physics:opacity}). \carla{} also assigns no calibrated reflectivity, so two physical quantities are set from measurements of real mirror phantoms and boards. The phantom reflectivity is the imaged surface's reflectivity multiplied by 0.14, the measured phantom-to-background ratio (0.14 to 0.17), in place of $\rho_m^2(\theta_m)\,\varepsilon(q,m)$ in \cref{eq:phantom_reflectivity}. The glint reflectivity is scaled by a gain set from the median glint ratio (\cref{eq:glint}) of the real boards (8.6). \Cref{fig:sensitivity}f and g vary both values.

The virtual sensor lies within $2.3 \times 10^{-5}$\,m of the exact mirror image at each tilt (details in Appendix~\ref{app:results}, \cref{tab:vsgeom}). To check the reflection law against a real mirror, we apply it to an indoor OS1-64 scene recorded without the mirror. With a real mirror in the same scene, the measured returns around the phantom lie a median of 6\,cm from the nearest simulated return. The same extension runs live with Autoware Universe~\cite{autoware2024universe} in the closed loop, where the simulated intersection provides the lanelet and point-cloud maps Autoware requires.\looseness=-1

\subsection{Datasets}
\label{sec:setup:datasets}
\label{sec:sim:scenes}

\begin{table}[!tb]
\caption{Evaluation datasets with drive and scan counts. Scored scans are the scans the metrics are computed on.}
\label{tab:datasets}
\centering
\tabfont
\setlength{\tabcolsep}{2pt}
\begin{tabular*}{\columnwidth}{@{\extracolsep{\fill}}l c c c c c@{}}
\toprule
dataset & what varies & drives & scans & scored & ground truth \\
\midrule
\multicolumn{6}{@{}l}{\textbf{\emph{simulation}}} \\
turn scenes & 3 tilts, 2 sources, 8 km/h & 18 & 3,303 & 296 & simulator \\
speed & 12 to 35 km/h & 53 & 4,338 & 268 & simulator \\
channels & 16 to 128 channels & 32 & 5,967 & 288 & simulator \\
reflective area & 0.36 to 3.50 m$^2$ & 30 & 5,560 & 679 & simulator \\
no attack & attack-off twins & 27 & 4,844 & 4,844 & all genuine \\
closed loop & 45\degree{}, 60\degree{}, 4 arms & 80 & live & -- & drive outcome \\
\midrule
\multicolumn{6}{@{}l}{\textbf{\emph{real recordings}}} \\
RW & 3 tilts $\times$ 3 areas & 9 & 2,302 & 224 & hand labels \\
public drives & city, no attack & 2 & 459 & 459 & all genuine \\
\bottomrule
\end{tabular*}

\end{table}

\Cref{tab:datasets} lists all evaluation data with drive counts, scored scans and the source of ground truth for each dataset.

\begin{figure}[!t]
\centering
\includegraphics[width=\columnwidth]{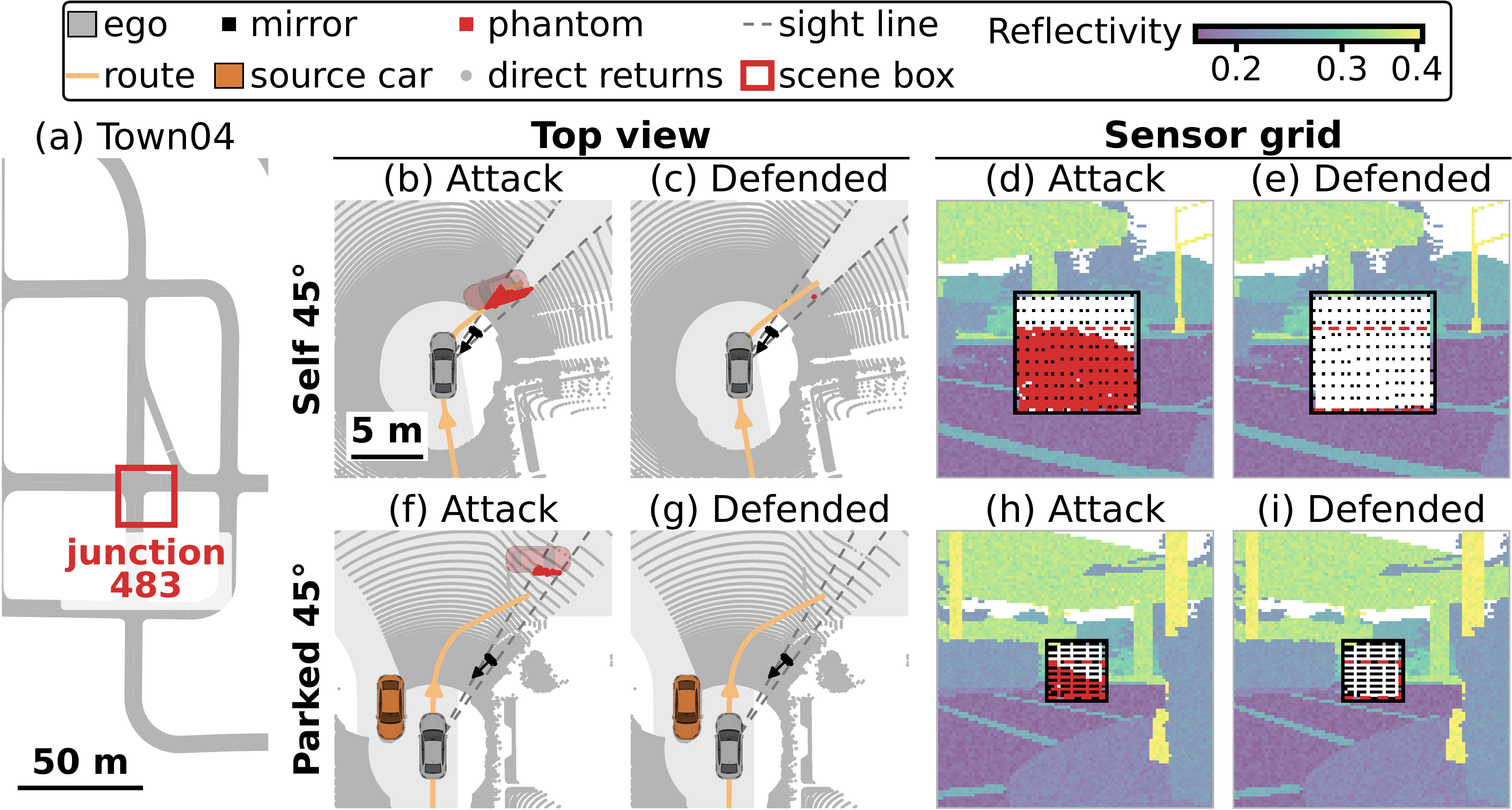}
\caption{Simulated self-image and parked-source attacks at 45\degree{}, per scene the scan with the most phantom returns from its source object. (a)~Town04. (b,c,f,g)~Top views, ego driving up, source object's image faint in~(b,f). (d,e,h,i)~Sensor grids, board boxed black, phantom boxed dashed red, reflectivity on the simulation's own log scale.}
\label{fig:simscans}
\end{figure}

\indent\textbf{Simulated Data.}~The simulated corpus is built at one junction of \carla{}'s Town04. \Cref{fig:simscans} shows one self-image and one parked-source scan at 45\degree{}, under attack and defended. Six scenes cover both scenarios of \cref{fig:threat} at the three tilts on an 8\,km/h turn, each with three repeats (spawn shifts of $-0.3$, $0$ and $+0.3$\,m along the lane). Three sweeps in \cref{tab:datasets} vary speed (12 to 35\,km/h, higher speeds on a straight route because the vehicle leaves the road on the turn at 20\,km/h), channels (16 to 128) and reflective area (0.36 to 3.50\,m$^2$). On the attack-free twins of \cref{sec:sim} every return \system{} removes is a false removal. A closed loop tests the attack and \system{} with a complete driving stack on the self-image scene, where the drive outcome is the ground truth. It runs at 45\degree{} and 60\degree{}, the tilts at which a board placed along its own plane at this junction both clears the braking corridor and keeps the self-image in view long enough. The corpus totals 133 attack drives in 36 sets, 19,168 attack scans and 80 closed-loop episodes (Appendix~\ref{app:results}), about 99\,GB.

\indent\textbf{Real-World Data.}~Nine drive-by recordings of the reference attack~\cite{yahia2025seeing} cover three tilts and three reflective areas (0.18, 0.36 and 0.60\,m$^2$). A single roof-mounted OS1-128~\cite{ouster2024os1} records at 10\,Hz while the vehicle approaches the board from about 20\,m at about 8\,km/h. We annotated a subset of scans per return in four classes (phantom-object, phantom-background, mirror, genuine-object) with a browser-based labeling tool we built,
because inside the board's footprint phantom-object, phantom-background and genuine returns lie side by side and no bounding box separates them. The annotator paints labels on individual returns in 3D and carries them from one scan to the next. Returns left unpainted carry no label. Each recording is scored on a fixed window of labeled scans, 224 across the nine recordings with 21,837 labeled phantom-object returns (\cref{fig:rwscans}).
Two public OS1-128 city-driving recordings without an attack (459 scans)~\cite{ouster2022samples} measure false removals in benign traffic.

\subsection{MirrorGuard Parameters}

\system{} runs with the same values in simulation and on the real sensor, except $T$ and $D_{\mathrm{FAST}}$, which the label-free startup rule (\cref{sec:design:constants}) sets per sensor from the first 30 warm-up and 30 calibration scans of one drive.
The rule gives $T=14.085$ and $D_{\mathrm{FAST}}=0.1091$ in simulation (a 6\,km/h drive without an attack) and $T=5.408$ and $D_{\mathrm{FAST}}=0.04875$ on the opening scans of the 45\degree{} 0.18\,m$^2$ recording.
We use this recording because no phantom object appears in its opening scans and its first phantom-object return comes 8.2\,s after the calibration window closes.
On both sensors the one-Lambertian floor of \cref{eq:kappat} binds on every candidate observation, so Path~D compares the candidate's median calibrated reflectivity with $D_{\mathrm{FAST}}$.
Of the confirmation lengths that the release hold of $\tau_{\mathrm{hold}}=0.5$\,s admits at the sensor rate, $N_{\mathrm{MIN}}=4$ confirms fastest.
The vehicle pose for the release comes from Autoware's localization in the closed loop, causal scan-to-map registration on the real recordings and the recorded pose in the other simulated drives.
\Cref{tab:notation} (Appendix~\ref{app:constants}) lists the remaining values.\looseness=-1

\subsection{Evaluation Metrics}
\label{sec:setup:metrics}

The metrics score \system{}'s removals against the ideal removal set of \cref{eq:ideal}, which holds every phantom-object return and no direct-path or glint return.
\textit{Object recall}, the primary metric, is the fraction of labeled phantom-object returns removed, pooled over the scored scans.
\textit{Whole-frame precision}, the headline precision, is the fraction of removed returns that are phantom returns and charges every non-phantom removal anywhere in the scan.
\textit{Path-only precision}, secondary, charges a non-phantom removal only on the driven corridor, 3.5\,m wide and 25\,m ahead, for returns more than 0.2\,m above the fitted road plane, while counting true positives over the whole scan.

On the real recordings, ground truth is the hand labels inside the scored windows, and unlabeled, mirror and genuine-object returns count as non-phantom.
In simulation, the origin label is the ground truth, and mirror, glint and native returns count as non-phantom.
Simulated recall counts the \textit{clearly appearing} (CA) scans, those with at least $n_{\mathrm{CA}}=100$ phantom-object returns, and scores both reflected images in the parked-source scenes.

\section{Evaluation}
\label{sec:eval}

The evaluation of \system{} starts in simulation, where every return has a known origin, and continues with the same design values on the real recordings of the reference attack.

\subsection{Simulation-Based Evaluation}
\label{sec:eval:simulation}
\label{sec:eval:carla}

The six simulated scenes test \system{} on both attack scenarios at the three tilts, with the origin of every return known exactly. \Cref{fig:simscans} shows one scan per scenario under attack and defended, and \cref{tab:carla} reports the results per scene.
Recall stays at or above 96.5\% and precision at or above 97.7\% in all six scenes.
Pooled over the 18 drives, \system{} removes 98.65\% of the phantom returns on the 296 clearly appearing scans at a precision of 98.88\%, and 98.37\% over all 517 scans that carry a phantom.
In every drive the first removal comes before the first clearly appearing scan, so \system{} is already removing the phantom before it becomes clearly visible in the scan.

Recall shows no trend with the tilt or the attack scenario.
The 45\degree{} self-image scene has the lowest recall, 96.5\%, and most of its missed returns fall in four scans taken with the vehicle 2.18 to 2.36\,m from the board.
When \system{} removes a non-phantom return, it is almost always a ground return right next to the phantom, and never a return of the board or of the parked car that sources the image.
In the 27 attack-free twins of \cref{tab:datasets} (4,844 scans), where no attack is present, \system{} removes no return at all.

\subsection{Transfer to Real-World LiDAR Measurements}
\label{sec:eval:realworld}
\label{sec:eval:results}
\label{sec:eval:rw}
\label{sec:eval:transfer}

\begin{figure}[!t]
\centering
\includegraphics[width=\columnwidth]{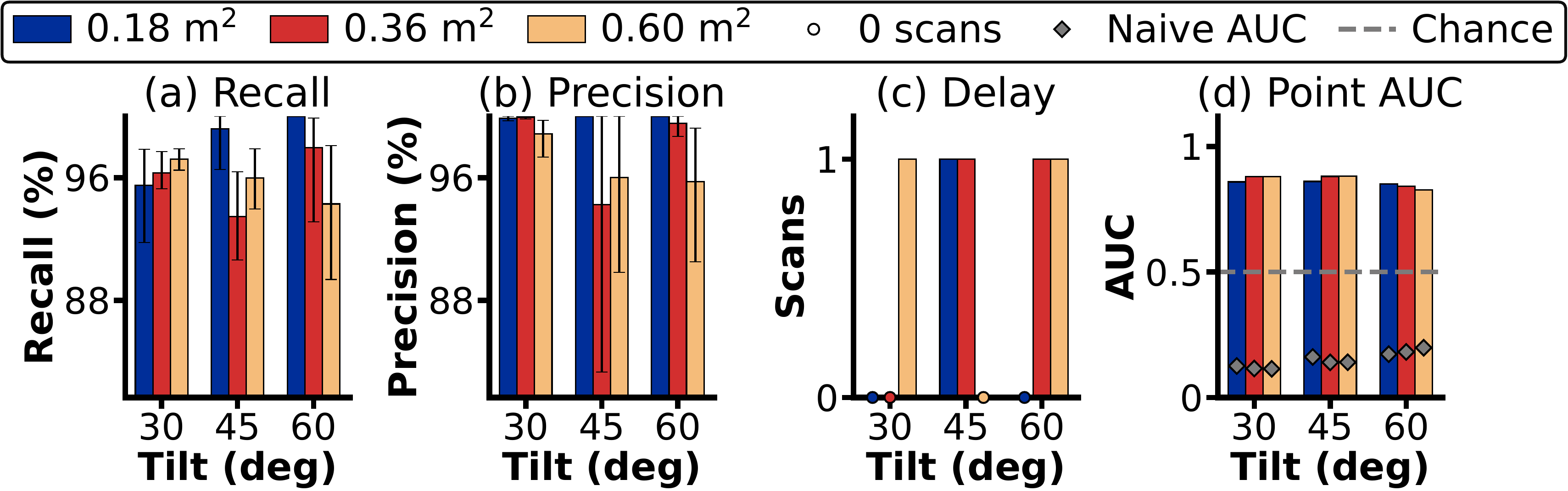}
\caption{RW point-level results per recording inside the scored window. Error bars are 95\% block-bootstrap intervals. (a)~Object recall. (b)~Whole-frame precision. (c)~Scans from the first phantom scan to the first removal. (d)~Point AUC of the darkness score against naive intensity.}
\label{fig:rw}
\end{figure}

\begin{figure}[!t]
\centering
\includegraphics[width=\columnwidth]{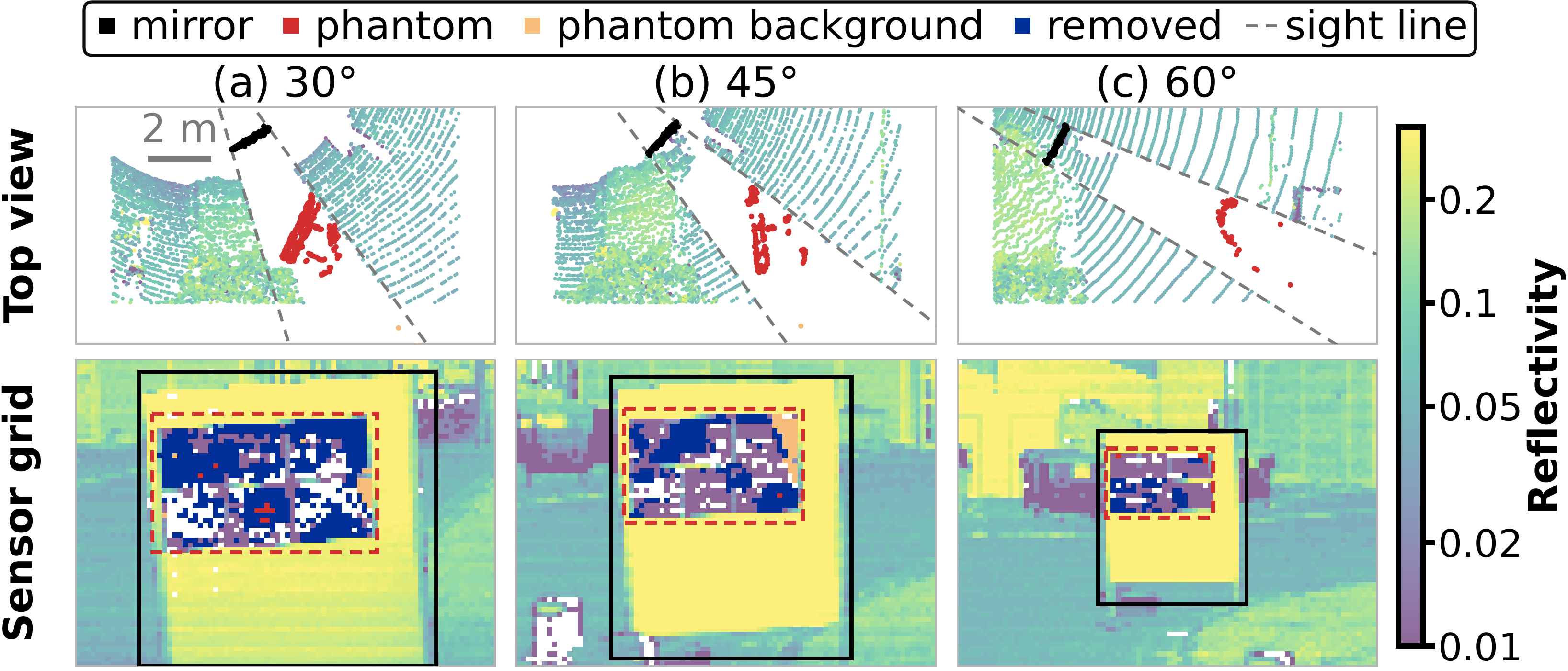}
\caption{One in-window scan of each RW recording with the 0.60\,m$^2$ board. Top: top view in the sensor frame, returns colored by calibrated reflectivity. Bottom: destaggered sensor grid, black outlines mark the mirror and red outlines the labeled phantom.}
\label{fig:rwscans}
\end{figure}

\begin{figure*}[!t]
\centering
\includegraphics[width=\textwidth]{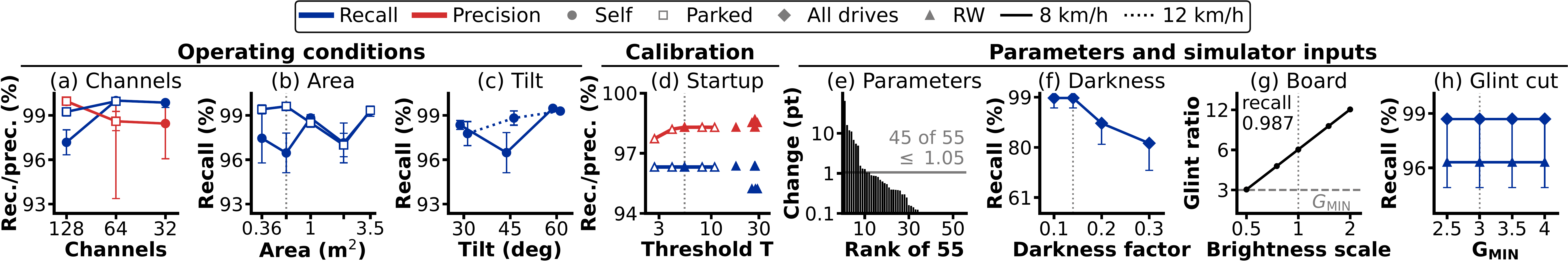}
\caption{Robustness of \system{}. Dotted verticals mark the applied value. Whiskers are SD over repeats in~(b,c), min to max in~(a) and mean to lowest in~(f,h). Open triangles in~(d) scale the calibrated pair from $\times$0.5 to $\times$2. (e)~Largest change per parameter.}
\label{fig:sensitivity}
\label{fig:carla}
\label{fig:robustness}
\end{figure*}

\system{} removes the phantom in all six simulated scenes, and we now test whether the same detector carries over to real LiDAR measurements, where the mirror board and its phantom are physical.
We run it on the nine real recordings of the reference attack with the design values used in simulation, left unchanged, and set only the startup pair $T$ and $D_{\mathrm{FAST}}$ on the real sensor, with the label-free rule of \cref{sec:design:constants} applied to its first scans.
With this pair, \system{} removes 96.31\% of the phantom-object returns of the nine recordings at 98.30\% whole-frame precision (\cref{fig:rw}a,b).
Even with the simulator's own pair, \system{} reaches the same 96.31\% recall at 98.30\% precision, so the result does not hinge on the exact pair the startup rule sets.
The simulator's radiometry takes two quantities from the real recordings, the phantom reflectivity ratio and the glint gain (\cref{sec:sim}), and beyond them this pair uses nothing from the real sensor.
Per recording, recall ranges from 93.47\% to 100.00\% (mean and SD 96.7\,$\pm$\,2.2\%) and whole-frame precision averages 98.2\,$\pm$\,2.3\%, so geometry and the phantom's darkness suffice on the nine recordings.

\begin{table}[t]
\caption{The six simulated scenes at 8\,km/h. CA: clearly appearing. Recall and precision: mean $\pm$ SD over three repeats. Onset to removal: from the first phantom return to the first removal, base repeat.}
\label{tab:carla}
\centering
\tabfont
\begin{tabular}{@{}llccccc@{}}
\toprule
tilt & scenario & CA & phantom & recall & precision & onset to \\
($^\circ$) & & scans & on lane & (\%) & (\%) & removal (s) \\
\midrule
30 & self-image & 16 & yes & 98.4$\pm$0.3 & 100.0$\pm$0.0 & 0.0 \\
30 & parked & 23 & no & 99.8$\pm$0.1 & 98.2$\pm$0.2 & 0.0 \\
45 & self-image & 17 & yes & 96.5$\pm$1.4 & 99.9$\pm$0.1 & 0.0 \\
45 & parked & 12 & yes & 99.2$\pm$0.2 & 100.0$\pm$0.0 & 0.6 \\
60 & self-image & 12 & yes & 99.5$\pm$0.1 & 98.9$\pm$0.1 & 0.0 \\
60 & parked & 18 & yes & 98.8$\pm$0.2 & 97.7$\pm$0.7 & 0.0 \\
\bottomrule
\end{tabular}

\end{table}

Beyond the phantom, the rest of the scene must stay in place.
Across all 22.9 million non-phantom returns in the scored scans, \system{} removes 483 and leaves 99.99\% untouched.
The region behind the detected boards holds the phantom, so wrong removals are most likely there.
In this region \system{} removes 48 of the 33,906 non-phantom returns (0.14\%).
A wrong removal would matter most on the driven path, and path-only precision there is 99.99\%.
In the two public drives a lead vehicle is visible in the audited scans, and \system{} removes none of the 64,006 object-height returns on the driven path (more details in Appendix~\ref{app:results}).

\Cref{fig:rwscans} shows one scan of each recording with the 0.60\,m$^2$ board. As in the simulated scans of \cref{fig:simscans}, the phantom lies beyond the board between the sight lines through it, and in these scans the removals stay inside the labeled phantom while the board and the scene around it remain in place.
The removal also begins early (\cref{fig:rw}c). On 4 of the 9 recordings the first removal falls on the first labeled phantom scan and on the other five one scan later, so the phantom stays untouched in at most one labeled phantom scan.
\Cref{sec:eval:robust} next changes the operating conditions, the calibration window and the defense parameters.\looseness=-1

\subsection{Robustness and Parameter Sensitivity}
\label{sec:eval:robust}

In simulation and on the real recordings, \system{} has so far run with one sensor setup and one parameter setting under the reference attack's conditions.
The sensor, the board and the speed are fixed in each RW recording, so the operating conditions vary in the simulated scenes.
In contrast, the startup calibration and the design parameters belong to \system{} itself and can be varied directly on the RW recordings.
With 128, 64 or 32 uniform channels and only the startup pair set anew for each, recall on the scored 45\degree{} scenes is 97.2 to 100\% at precision of at least 98.4\% (\cref{fig:sensitivity}a).
A deployed \system{} thus moves to a sensor with a quarter of the RW sensor's channels without a design change.
From a small 0.36\,m$^2$ board to a large 3.50\,m$^2$ one, \system{} keeps a mean recall of 96.5 to 99.6\% (\cref{fig:sensitivity}b).
On the self-image scene, the 8\,km/h turns of \cref{tab:carla} keep mean recall between 96.5 and 99.5\% across the tilts (\cref{fig:sensitivity}c).
A faster turn at 12\,km/h keeps mean recall at or above 97.8\% at every tilt.\looseness=-1

The startup rule runs once per sensor on the opening scans of one drive, so we test whether the choice of that drive matters.
Calibrated on the opening scans of each of the nine RW recordings in turn, \system{} keeps recall between 95.23 and 96.36\% at whole-frame precision between 98.30 and 98.69\% (\cref{fig:sensitivity}d).
Scaling $T$ and $D_{\mathrm{FAST}}$ each from half to twice their calibrated values keeps recall at 96.31\% at whole-frame precision between 97.73 and 98.32\% (\cref{fig:sensitivity}d), and \cref{tab:robust} in Appendix~\ref{app:results} lists the nine windows and the worst of the 25 pairs.
A sensor thus calibrates in the first six seconds of one drive, and an imprecise calibration barely changes the outcome.

Beyond the startup pair, each of the 55 design parameters moves alone from half to twice its value (\cref{fig:sensitivity}e).
Wherever the startup rule calibrates, 45 of them change neither RW metric by more than 1.05 points.
The other ten each shape a geometric or radiometric test, so halving or doubling one of them changes the test itself and gives its largest change.
Even so, no single-parameter setting removes a return on the 27 attack-free twins.
No design value was picked from this sweep, and 11 of its 208 single-parameter settings even raise recall by up to 0.34 points and precision by up to 1.28 points.
A deployed \system{} still takes the design values as published and calibrates only its startup pair.

The next test returns to the simulator and moves its two radiometric inputs (\cref{sec:sim}).
Raising the simulated ratio from 0.14 to 0.30 makes the phantom about twice as bright and lowers mean recall from 98.7 to 81.6\% (\cref{fig:sensitivity}f).
In every drive, removal still starts on the same scan as at 0.14.
A dimmer mirror lowers $\rho_m$ in \cref{eq:phantom_reflectivity} and only makes the phantom darker.
At half the glint gain the simulated board still clears the glint threshold $G_{\mathrm{MIN}}$, so mean recall stays at 98.7\% (\cref{fig:sensitivity}g).
On the RW recordings, every tested $G_{\mathrm{MIN}}$ from 2.5 to 4.0 gives the same 96.31\% recall (\cref{fig:sensitivity}h), and other structures also clear the glint gate, which only proposes candidates.
In deployment, precision thus rests on opacity and darkness, not on the glint gate.

A faster vehicle sees the phantom in fewer scans, so the last test drives a simulated straight route at 20 to 35\,km/h.
At 20\,km/h every drive shows the phantom in 5 to 8 scans and at 30\,km/h in 4 to 5, so every drive reaches $N_{\mathrm{MIN}}$.
At both speeds removal starts on the first phantom scan, and recall is at least 99.5\% at 20\,km/h and 99.7\% at 30\,km/h.
At 35\,km/h \system{} removes the phantom on 4 of 5 drives per tilt at 45\degree{} and 60\degree{}, and each missed drive shows it in only 3 scans, fewer than $N_{\mathrm{MIN}}$.
On these two drives Autoware's tracker publishes the phantom in at most one message.
\system{} thus covers the typical intersection approach speed of up to 20\,km/h~\cite{yahia2025seeing} with a margin.
\Cref{sec:eval:baselines} next compares \system{} with detectors trained on the labeled recordings.

\label{sec:eval:clean}

\subsection{Comparison with Learning-Based Baselines}
\label{sec:eval:baselines}

\begin{table}[!tb]
\caption{\system{} against five learned detectors, mean $\pm$ sample SD over three seeds. Trained on sim.\ gives RW recall/precision (\%) after simulation-only training, Unseen tilt on the 30\degree{} and 60\degree{} recordings with the test tilt left out of training. Latency is CPU/GPU (8 threads, one L40S). \system{}, sim.\ is the per-scene range of \cref{tab:carla}.}
\label{tab:learned}
\centering
\tabfont
\setlength{\tabcolsep}{4pt}
\begin{tabular}{@{}lccccc@{}}
\toprule
Method & Recall & Whole-frame & Trained & Unseen & Latency \\
 & (\%) & prec. (\%) & on sim. & tilt & (ms) \\
\midrule
\multicolumn{6}{@{}l}{\textit{\textbf{Physics-based, no learned model}}} \\
MirrorGuard, sim. & 96.5 to 99.8 & 97.7 to 100.0 & -- & -- & -- \\
MirrorGuard, RW & 96.31 & 98.30 & -- & 96.6/99.0 & 59 \\
\midrule
\multicolumn{6}{@{}l}{\textit{\textbf{Trained on RW labels, one recording left out}}} \\
LightGBM~\cite{sevgen2023lightgbm} & 55.83$\pm$1.22 & 34.75$\pm$1.65 & 21.7/0.9 & 9.3/7.6 & 196 \\
PointNet~\cite{qi2017pointnet} & 47.78$\pm$3.57 & 43.79$\pm$0.77 & 32.1/0.7 & 5.5/5.4 & 618 \\
WeatherNet~\cite{heinzler2020cnn} & 70.19$\pm$6.08 & 62.52$\pm$2.98 & 0.1/1.5 & 23.9/43.5 & 664/23 \\
SalsaNext~\cite{cortinhal2020salsanext} & 88.77$\pm$14.03 & 99.21$\pm$0.62 & 56.1/0.9 & 60.9/91.1 & 287/19 \\
LMNet~\cite{chen2021moving} & 98.38$\pm$0.26 & 97.22$\pm$2.75 & 3.1/2.9 & 50.0/97.9 & 523/154 \\
\bottomrule
\end{tabular}

\end{table}

Five published detectors learn from labels what \system{} derives from the physics of reflection (\cref{tab:learned}).
Each trains on seven RW recordings, sets its threshold on an eighth and is scored on the ninth.
The shared site, boards and phantom favor them, yet three stay far below \system{} in mean recall and precision.
LMNet reaches a higher mean recall at a lower mean precision and SalsaNext the reverse.
In both, the weaker metric moves with the training seed (per-seed values in Appendix~\ref{app:learned}), whereas \system{} is deterministic and gives the same result on every run.
Their training always includes the test tilt.
Without it, both keep their recall on a tilt between their training tilts but fall to 50.0\% and 60.9\% mean recall outside them (\cref{tab:learned}).
Trained on simulation, all five find at least 96.6\% of the phantom returns on unseen simulated drives but remove far more non-phantom than phantom returns on RW, while \system{} runs on both with the same design values.

Laser-injection defenses seek returns that break physical consistency or are too bright, but a mirror return travels a real optical path and needs its own defense.
CARLO~\cite{sun2020towards} on a tight box and ADoPT~\cite{cho2023adopt} miss all 191 RW phantoms, and Shadow-Catcher~\cite{hau2021shadowcatcher} does not separate them from genuine vehicles.
Per-point brightness~\cite{park2024intensity} ranks the phantom below chance (AUC~0.132 against 0.875 for the darkness score, \cref{fig:rw}d).\looseness=-1

\indent\textbf{Computational Cost.} \system{} needs no GPU. On an Intel Xeon Gold 5420+, the complete filter with release takes 59\,ms per scan on average over the nine RW recordings, and 98.9\% of scans finish within the sensor's 100\,ms period (per-thread times in Appendix~\ref{app:results}). On the same CPU, the mean time per scan of every learned detector of \cref{tab:learned} exceeds this period, and the four profiled ones execute 2.5 to 9.1 times more instructions. The in-vehicle computers tested with Autoware already include an eight-core Xeon~\cite{autowareadcomputers,intelxeone2278ge}. \Cref{sec:eval:perception} traces the filtered cloud into perception.

\subsection{Impact on Downstream Perception}
\label{sec:eval:perception}
\label{sec:eval:autoware}

\begin{figure}[!tb]
\centering
\includegraphics[width=\columnwidth]{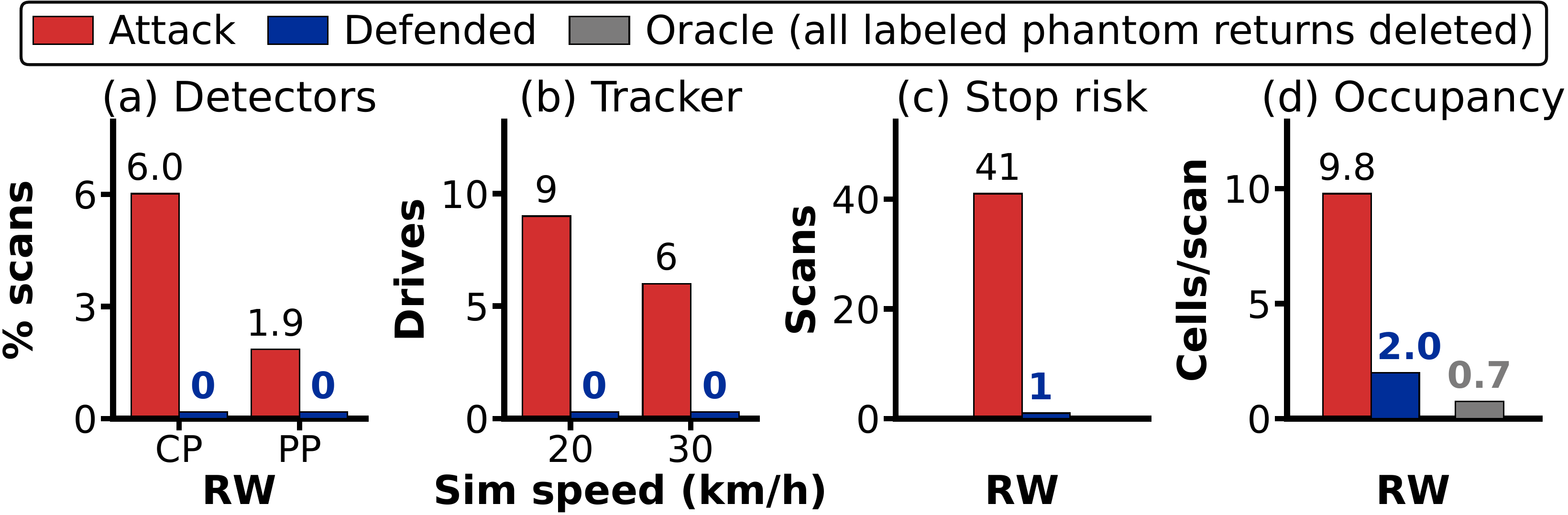}
\caption{Downstream effect of \system{}. (a)~RW phantom scans with a CAR within 3\,m of the phantom. (b)~Simulated drives with a published phantom CAR, of 10 per speed. (c)~Stop-risk scans of the three 0.60\,m$^2$ RW recordings. (d)~Occupied grid cells per scan.}
\label{fig:downstream}
\end{figure}

\system{} removes the phantom at the point level, both in simulation and on the RW recordings.
We now test whether the phantom also disappears from the perception of Autoware~\cite{kato2018autoware,autoware2024universe}, while real objects remain.
Its detectors CenterPoint~\cite{yin2021centerpoint} and PointPillars~\cite{lang2019pointpillars} place boxes on objects, a tracker follows obstacles over time, and an occupancy grid marks occupied cells.
\system{} filters the point cloud before all three, so none of them is changed or retrained.

Even from a single scan, CenterPoint labels the phantom a CAR on all three recordings with the 0.60\,m$^2$ board.
Its highest score of 0.705 matches the median score of the real objects it detects (0.70).
\Cref{fig:downstream}a counts all 216 RW scans with a phantom, including those of the two smaller boards.
At a score gate of 0.35, CenterPoint labels the phantom a CAR on 13 of them and PointPillars on 4, and with \system{} neither detector does.
On the simulated straight-route drives, Autoware's shipped tracker reports the phantom as a CAR on most drives and on none with \system{} (\cref{fig:downstream}b).

Autoware's planner stops for any obstacle on the vehicle's path, not only for a CAR.
On the RW recordings, Euclidean clustering turns the phantom returns into such obstacles for the tracker.
The recordings have no lane map or route, so we count a scan as a stop risk when a phantom track overlaps the vehicle's swept path plus the planner's 0.3\,m stop margin.
On the three recordings with the 0.60\,m$^2$ board, this happens on 41 of the 85 scans, and on 1 with \system{} (\cref{fig:downstream}c).
In the occupancy grid, the phantom fills 9.8 cells per scan.
With \system{} 2.0 cells remain, and 0.7 remain with an oracle that deletes every labeled phantom return (\cref{fig:downstream}d).
MirrorGuard keeps the real objects in view. Of the 3,296 real objects that CenterPoint detects without filtering, it still detects 3,275 after filtering.
\Cref{sec:eval:closedloop} puts the planner in the loop to test whether the vehicle still stops.

\subsection{Closed-Loop Driving Evaluation}
\label{sec:eval:closedloop}

\begin{figure*}[!t]
\centering
\includegraphics[width=\textwidth]{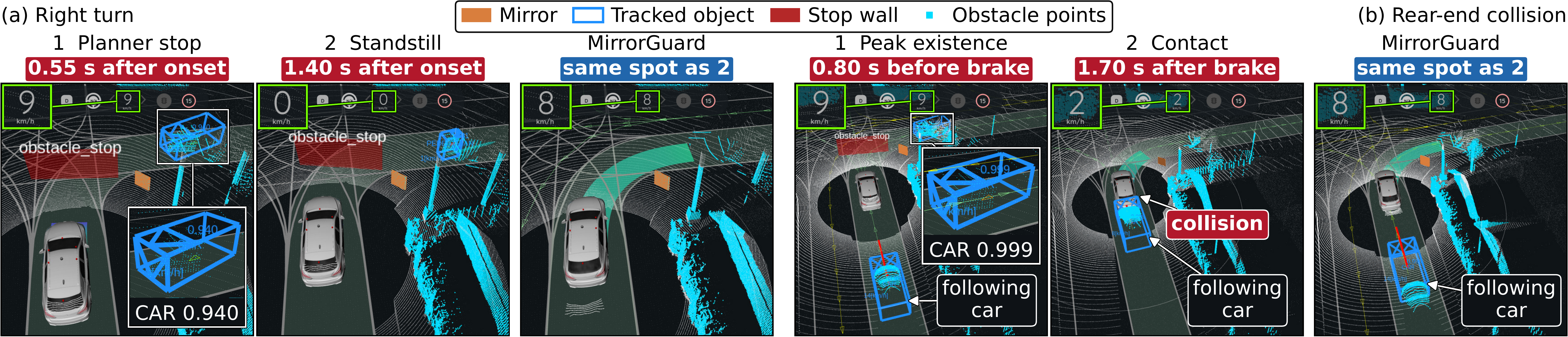}
\caption{Autoware at the same road position, one episode per scenario. White bubbles magnify the phantom's tracked object with its class and existence probability, green boxes the speed display. (a)~Right turn. In panel 2 the CAR track is already deleted, and the box is a short new track that Autoware displays as a PEDESTRIAN. (b)~Rear-end collision.}
\label{fig:autoware}
\end{figure*}

\begin{table}[!tb]
\caption{Closed-loop outcome per arm and tilt. Stop counts the episodes with a planner stop, Min.\ speed is the median minimum speed in m/s, and Rear-end counts the crashes of the following car at 45\degree{}.}
\label{tab:closedloop}
\centering
\tabfont
\begin{tabular}{@{}lccccc@{}}
\toprule
 & \multicolumn{2}{c}{45\degree{}} & \multicolumn{2}{c}{60\degree{}} & \\
\cmidrule(lr){2-3}\cmidrule(lr){4-5}
Arm & Stop & Min.\ speed & Stop & Min.\ speed & Rear-end \\
\midrule
Clean, \system{} & 0/10 & 1.92 & 0/10 & 2.00 & 0/5 \\
Board, no reflection & 0/10 & 1.98 & 0/10 & 2.05 & 0/4 \\
Attack & 10/10 & 0.00 & 10/10 & 0.00 & 5/5 \\
Attack, \system{} & 0/10 & 2.00 & 0/10 & 2.01 & 0/5 \\
\bottomrule
\end{tabular}

\end{table}

\begin{figure}[!tb]
\centering
\includegraphics[width=\columnwidth]{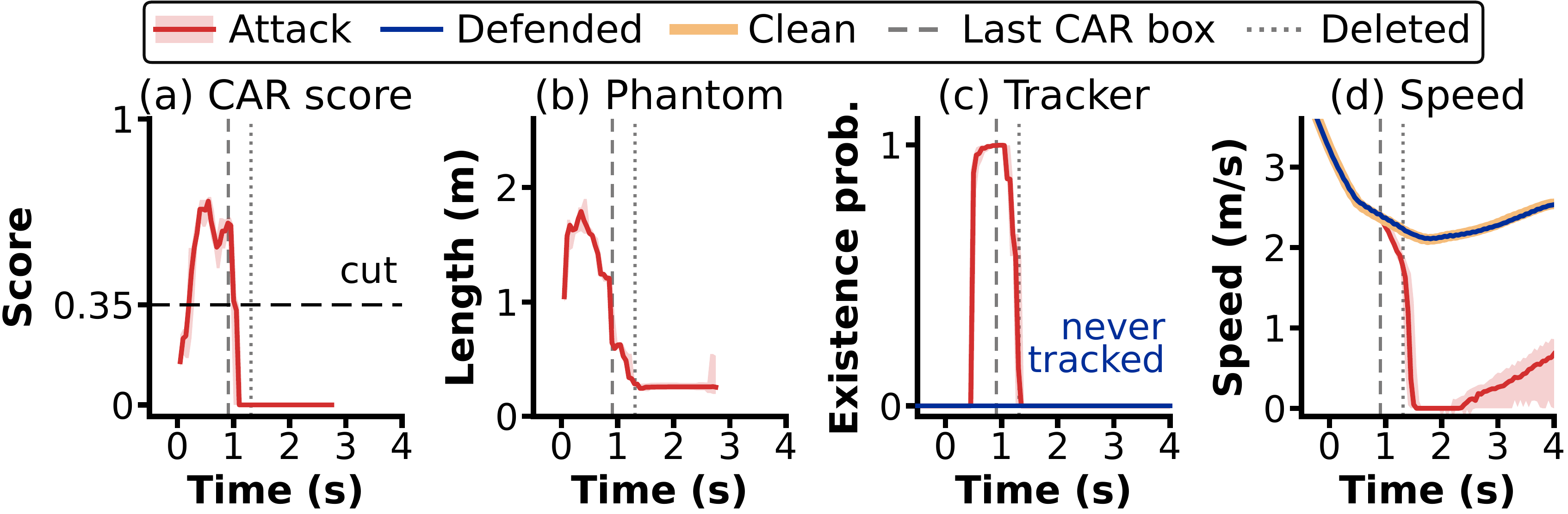}
\caption{Closed-loop mechanism at 45\degree{}, median and range over 10 episodes aligned at the median onset ($t{=}0$). (a,b)~Frames with phantom returns in at least 8 episodes. (a)~CenterPoint CAR score. (b)~Phantom length along the driving direction. (c)~Track existence. (d)~Ego speed.}
\label{fig:clmech}
\end{figure}

In the closed loop, Autoware drives a vehicle in CARLA through the self-image scene at 45\degree{} and 60\degree{}.
CenterPoint feeds Autoware's shipped tracker and planner.
Each tilt has the four arms of \cref{tab:closedloop}, with ten episodes each.
Within an arm, the episodes differ only in the stack's start-up timing.

The planner stops in every attack episode and in none of the others (\cref{tab:closedloop}).
At 45\degree{} the board with its reflection off produces no phantom return, object or stop, so the reflection causes the stop.
At this tilt all 88 stop messages of the attack episodes name a phantom track.
With \system{} no episode stops, and the vehicle drives as fast as in the clean drive.
In these episodes \system{} removes 98.89\% of the phantom returns at 45\degree{} and 98.36\% at 60\degree{}.
The 22 to 40 phantom returns left per episode at 45\degree{} form no track.

At 45\degree{} every attack episode follows the same course (\cref{fig:autoware}a, \cref{fig:clmech}).
CenterPoint labels the phantom a CAR, and the tracker raises the existence probability of its track to near 1.
As the vehicle closes in, the visible phantom shrinks and its CAR score falls below 0.35.
The tracker then deletes the CAR track, while 104 to 278 phantom returns still arrive per scan.
The vehicle halts for a median 0.88\,s and moves off, and no halt at either tilt lasts 2\,s.

A short halt with no visible cause endangers the traffic behind, so further episodes at 45\degree{} add a following car to each arm (\cref{fig:autoware}b).
The follower car uses a standard car-following model~\cite{treiber2000congested,treiber2006delays} and reacts to the vehicle's braking after 1.75\,s, within the reaction times measured for unexpected events~\cite{green2000perception}.
The follower rear-ends the vehicle in every attack episode and in no other, and under attack with \system{} it never comes closer than its 2.0\,m standstill gap.

\begin{table}[t]
\centering
\caption{Ablation pooled over the nine RW recordings and 18 simulated drives. Each column removes one part of the detector, and No radio.\ removes every darkness test. R/P are phantom-object recall and precision (\%), whole-frame precision on RW and recall on clearly appearing simulated scans.}
\label{tab:ablation1}
\tabfont
\setlength{\tabcolsep}{2.5pt}
\begin{tabular}{@{}lccccccc@{}}
\toprule
& \multicolumn{2}{c}{Physical constraints} & & \multicolumn{4}{c}{Mechanism ablations} \\
\cmidrule(lr){2-3}\cmidrule(lr){5-8}
R/P & No opac. & No radio. & \textbf{Full} & No track & No latch & No comp. & No alarms \\
\midrule
Sim. & 98.7/98.9 & 79.6/98.6 & \textbf{98.7/98.9} & 85.7/99.4 & 98.7/98.9 & 98.6/99.4 & 98.4/99.1 \\
RW & 96.3/95.6 & 89.2/97.8 & \textbf{96.3/98.3} & 94.1/98.7 & 96.3/96.6 & 95.2/98.6 & 94.1/98.3 \\
\bottomrule
\end{tabular}
\end{table}

In all 25 clean drives with \system{} no return is removed or held, so the planner receives the same cloud as without the defense.
In the attack episodes with \system{}, returns outside the held set of \cref{sec:design:output} reach Autoware 0.13\,s after their scan, and held returns 0.73\,s after it. %
During these 0.73\,s the vehicle travels 4.1\,m at 20\,km/h and 6.1\,m at 30\,km/h.
About 95\% of a real scan is released at once, because on the RW recordings the held set averages 4.9\% of a scan.
Without the hold, the held returns would be released one scan period after their scan, but recall on the RW recordings would fall from 96.31 to 94.92\%.
\Cref{sec:eval:ablation} next takes the mechanisms of \system{} out one at a time.
\subsection{Contribution of Defense Mechanisms}
\label{sec:eval:ablation}

To measure what each part of \system{} contributes, we remove it from the full defense and keep the others (\cref{tab:ablation1}).
Without opacity, \system{} no longer rejects candidates whose returns do not see through the board.
Recall is unchanged, but RW whole-frame precision falls from 98.3 to 95.6\%, and the simulated results do not change.
Radiometric evidence adds recall.
It measures how dark a structure is against its background, and the phantom is dark (\cref{fig:rw}d).
Without radiometric evidence, recall falls from 96.3 to 89.2\% on the RW recordings and from 98.7 to 79.6\% in simulation, while precision changes little (97.8 and 98.6\%).

The continuation tracker and the bracketed carry keep removing the phantom on the scans where no core is confirmed.
Without them, recall falls to 94.1\% on the RW recordings and to 85.7\% in simulation.
In simulation they also account for most of the non-phantom returns that \system{} removes, so precision rises from 98.9 to 99.4\% without them.
The track latch and core completion add smaller shares.
Without the latch, RW whole-frame precision falls to 96.6\%, and without completion, RW recall falls to 95.2\%.
Without the micro alarm and the sparse crossing, recall is 94.1\% on the RW recordings and 98.4\% in simulation.
\Cref{fig:ablation} in Appendix~\ref{app:results} shows the mechanism ablations per recording and the
aperture test alone, which reaches only 58.40\% whole-frame precision on the RW recordings.
The next subsection turns to an attacker who knows these parts.

\subsection{Adaptive Attack and Security Analysis}
\label{sec:security}

We consider a white-box adversary with full knowledge of \system{}'s design and parameters. We evaluate two adaptive attacks that seek to evade radiometric confirmation through phantom brightening or induce the removal of legitimate returns by placing the phantom near a real object.

The darkness test relies on a phantom that is darker than the background at the same range, and the recorded RW phantoms reach only 0.14 to 0.17 times this background. In the parked-source scenario the attacker chooses the source object and can look for a brighter one. A car that looks bright to the eye can still be dark to the LiDAR. One RW recording includes a white car parked at the roadside, and when the board reflects this car its phantom reaches only 0.135 times the background. We therefore raise the brightness of the phantom returns directly, in the 18 simulated drives, on the labeled RW phantoms by scaling existing returns without adding new ones, and in the closed-loop drives, which keep their scene and change only the phantom brightness. At about twice the recorded brightness, the darkness test still confirms all 18 simulated drives at the same first scan as at the recorded brightness. Recall is 81.6\% on average in simulation and 89.30\% on the RW recordings. In the closed loop the undefended vehicle stops for the phantom. With \system{} it does not stop in any of the 10 drives and keeps a minimum speed of 2.03\,m/s, because \system{} removes 98.63\% of the phantom returns and Autoware forms no phantom track. A deployment that also has to cover brighter phantoms runs \system{} without radiometric evidence (No radio.\ in \cref{tab:ablation1}). This configuration confirms the phantom from aperture, opacity and persistence alone. Its recall stays at 89.2\% on the RW recordings from the recorded brightness to 100 times brighter and at 79.6\% in simulation to 14 times brighter, at 97.8 and 98.6\% precision. In the closed loop at about 3.6 times the recorded brightness, the vehicle with this configuration does not stop in any of the 10 drives.

In the second attack, a real object stands next to the phantom in the simulated parked-source drives. Pooled over the tilts, \system{} removes at most 7.2\% of a pedestrian's returns and 1.2\% of a car's. A car darkened to the phantom's brightness keeps all of its returns from a gap of 3\,m.

The attacker can also act on the board itself. Its area and its glint are varied in simulation (\cref{fig:sensitivity}b and~g), and on the RW recordings we emulate a less visible board. With 75\% of the recorded board returns deleted, recall stays at 88.97\%, and with the board returns dimmed to half it stays at 96.61\%.\looseness=-1

\section{Discussion}
\label{sec:discussion}

\system{} is built on the physical constraints of the mirror reflection. It is a point-level filter that reads the range, direction and reflectivity a time-of-flight LiDAR reports for each return. It runs before any perception module and is not part of Autoware or of another driving stack, so one filter protects different detection and planning architectures without changes to their components. On a new sensor, the startup rule re-derives $T$ and $D_{\mathrm{FAST}}$ without labels, and the two design rules re-run for its beam geometry. In simulation, setting only $T$ and $D_{\mathrm{FAST}}$ anew keeps recall of at least 97.2\% at precision of at least 98.4\% on the scored scenes with 128, 64 or 32 channels. We evaluate this design in simulation across tilts, board areas, speeds, channel counts and both scenarios, and on the nine RW recordings of the reference attack. The simulator's reflection arithmetic is exact and its radiometry is a model, so simulation results inherit this model. Intermediate tilts, other physical sensors and other sites lie outside the evaluated scope.\looseness=-1
\label{sec:limitations}

The physical constraints behind \system{} hold for static, planar mirrors under single-bounce reflection. Relays through a second specular surface and curved or moving reflectors require extending the geometric and temporal model. Everyday reflective surfaces such as glazed facades and wet roads reflect beams in ordinary driving, and separating these benign mirror images from attacks is an open research question. On the public city drives the vehicle passed a storefront line and reflective street furniture and \system{} removed nothing on the driven corridor, and a dedicated test of glazed facades and wet roads remains open.

Beyond the reflector itself, three directions remain open for future work. A multi-return or frequency-modulated continuous-wave (FMCW) LiDAR may observe the mirror and its image in one measurement, which would make the opacity and darkness evidence direct. Rain and fog can alter mirror reflectivity and beam propagation~\cite{yahia2025seeing}, and sunset or night change the ambient light, so the attack and \system{} should next be measured together across weather and times of day. Finally, repeating the closed-loop drives on a vehicle at a test track would take this evaluation to the road.

\system{} reads only the stream the sensor already ships, so deploying it on a vehicle is a software change. The main deployment trade-off is temporal confirmation, which must balance reliable phantom removal against the timely delivery of sensor measurements. Selective buffering holds only the returns under suspicion, so most of the point cloud is delivered without this delay.

\section{Related Work}
\label{sec:related}

\indent\textbf{Attacks with Physical Objects.} Attacks that emit nothing present no foreign signal for injection defenses to detect. Wan et~al.~\cite{wan2022tooafraid} showed that benign objects such as cardboard boxes off the road stop conservative planners permanently. Adversarial objects fool learned detectors~\cite{cao2019objects,tu2020physically,zhu2021arbitrary,yang2021roadside,zhu2024aemorpher,cao2021invisible}. These objects are computed against a learned detection model, and the defenses against them harden that model. A mirror needs no knowledge of the detector because it forges the phantom's position in the raw measurements, so the phantom reaches every consumer of the point cloud.

\indent\textbf{Reflector-Based LiDAR Attacks.} Shadow Hack~\cite{kobayashi2024shadowhack,kobayashi2025invisible} removes ground returns with reflective or absorptive mats so a detector reports a false car. MirrorDrift~\cite{nagata2026mirrordrift} actuates the reflector to corrupt localization on a sensor with interference mitigation. Yahia et~al.~\cite{yahia2025seeing} showed that a planar mirror both erases and adds obstacles. The same study concluded that a countermeasure must reason about \textit{path plausibility}, whether the spatial configuration of the observed returns is consistent with single-bounce propagation. \system{} is such a test.

\indent\textbf{Mirror Detection and Glare Suppression.} Outside the adversarial setting, Yang and Wang~\cite{yang2011solving} detect mirrors in planar scans, Koch et~al.~\cite{koch2017detection} and Zhao et~al.~\cite{zhao2019reflection} detect reflected points in 3D scans and remove them or mirror them back, and others handle glass and mirror-like surfaces in robot navigation~\cite{tibebu2021lidar,damodaran2023mirror}. Other methods need more than a stock point cloud, such as the multibounce measurements of Henley et~al.~\cite{henley2023multibounce}, the full waveforms of Ghost-FWL~\cite{ikeda2026ghostfwl} or low-level echoes before the point cloud forms~\cite{gump2026glare}. \system{} inherits one observation from Yang and Wang~\cite{yang2011solving}, that the reflected world lies behind the mirror plane and inside the mirror's outline, but here an adversary places the mirror.

\section{Conclusion}
\label{sec:conclusion}

We presented \system{}, to our knowledge the first software-only, point-level defense that removes the phantom returns of a physical passive-mirror attack on a vehicle-mounted LiDAR. On the nine RW recordings it removes 96.31\% of the phantom vehicle's returns at 98.30\% precision and leaves 99.99\% of the 22.9 million non-phantom returns untouched.
The simulator's confirmation thresholds applied unchanged reproduce these numbers on the real sensor. %
With \system{} no closed-loop reference-attack drive produces a phantom stop or a rear-end collision, and Autoware's tracker with Euclidean clustering holds the phantom on the path in 41 of 85 scans without the defense and in 1 with it. %
\system{} costs 59\,ms mean per scan on a CPU, and the four profiled learned baselines execute 2.5 to 9.1 times more instructions. %

\bibliographystyle{IEEEtran}
\bibliography{refs}

\makeatletter\begingroup\let\@elt\relax\edef\@tempa{\@botlist\@deferlist\@dbldeferlist}\ifx\@tempa\@empty\endgroup\else\endgroup\clearpage\fi\makeatother
\appendices
\suppressfloats[t]
\section{Ethics Considerations}
\label{app:ethics}

This paper is a defense against a published attack~\cite{yahia2025seeing} and introduces no new attack capability. The reference attack is public, and its authors state that all their experiments ran in a controlled environment with no risk to public safety and that they disclosed the vulnerability to support the development of reflection-aware perception and validation systems~\cite{yahia2025seeing}. Our Autoware and closed-loop results reproduce the published attack's consequences in simulation and show how they arise. They add no attack path that the reference work did not already demonstrate, so no vendor disclosure is pending. The real attack recordings were made on a closed campus site with a safety driver and obtained from its authors~\cite{yahia2025seeing}. We do not release them publicly. Our own experiments are simulation and offline replay only, and no road test was conducted. A wrong removal from a perception stream is safety-relevant, so we measure real returns removed and real detections lost as first-class results (\cref{sec:eval:results,sec:eval:autoware,sec:eval:closedloop}).

We publish the defense's constants and analyze the attacker who knows them (\cref{sec:security}). The work has no human participants and no collected personal data. The attack recordings come from a closed test site, and the attack-free recordings are publicly distributed Ouster OS1-128 sample drives~\cite{ouster2022samples}, used for evaluation only, in which people appear only as unlabeled points that we neither label nor try to identify. The simulation extension models the attack's mirror physics to test defenses without road experiments. Because the work involves no human participants and no personal data, it was not submitted for ethics-board review.

\section{Constants and Their Provenance}
\label{app:constants}

\begin{table}[!tb]
\centering
\caption{Constants of \system{}. S~=~simulated drive without an attack. F~=~opening scans of the field sensor.}
\label{tab:notation}
\label{tab:constants}
\label{tab:structural}
\label{tab:allconstants}
\label{sec:setup:constants}
\tabfont
\begin{tabular}{@{}l l@{\hspace{10pt}}l l@{}}
\toprule
\multicolumn{4}{@{}l@{}}{\textit{Startup rule, per sensor}} \\
$T^{S}$ & 14.085 & $T^{F}$ & 5.408 \\
$D_{\mathrm{FAST}}^{S}$ & 0.1091 & $D_{\mathrm{FAST}}^{F}$ & 0.04875 \\
\multicolumn{4}{@{}l@{}}{\textit{Design rule}} \\
$G_{\mathrm{MIN}}$ & 3.0 & $D_{\mathrm{IMG}}$ & 19.0 m \\
$N_{\mathrm{MIN}}$ & 4 observations &  &  \\
\multicolumn{4}{@{}l@{}}{\textit{Stated rule in simulation}} \\
\multicolumn{4}{@{}l@{}}{Sparse threshold 0.5245 (1.10$\times$ the largest simulated track score without an attack)} \\
\multicolumn{4}{@{}l@{}}{\textit{Micro alarm: set on the characteristics of~\cite{yahia2025seeing}}} \\
$N_V$ & 2 observations & $t_V$ & 0.10 s \\
$\rho_V$ & 0.50 & $\omega_V$ & 0.80 \\
$r_V$ & 20 m & $d_{\mathrm{lo}},d_{\mathrm{hi}}$ & 2, 8 m \\
$\theta_V$ & 8$^{\circ}$ & $\gamma_{\mathrm{lo}},\gamma_{\mathrm{hi}}$ & 1.30, 2.25 \\
$h_V$ & 0.30 m &  &  \\
\multicolumn{4}{@{}l@{}}{\textit{Sensor}} \\
$\mu$ & 0.18$^{\circ}$ & $I_{\mathrm{floor}}$ & 1 Lambertian unit \\
\multicolumn{4}{@{}l@{}}{\textit{Fixed design parameters}} \\
$n_{M,\min}$ & 80 returns & $\sigma_{\max}$ & 0.12 m \\
$\nu_{\max}$ & 0.30 & $r_{M,\min},r_{M,\max}$ & 2, 18 m \\
$k_F$ & 3 (second family 2) & $\omega_{\min},\phi_{\min}$ & 0.5, 0.5 \\
$f_{\mathrm{band}}$ & 0.5 & $H_b$ & [0.2, 2.5] m \\
$f_{\mathrm{supp}}$ & 0.30 & $\Delta r$ & 2 m \\
$n_{B,\min}$ & 20 returns & $N_{\mathrm{cap}}$ & 30 observations \\
$r_{\mathrm{floor}}$ & 1 m & $\kappa_0$ & 0.25 \\
$N_P$ & 4 observations & $\tau_{\mathrm{hold}}$ & 0.5 s \\
$w_c$ & 8 cells & $R_c$ & 0.6 m \\
$b$ & 0.10 m & $\alpha_c$ & 3 \\
$W_{\mathrm{carry}}$ & 32 triples & $r_{\mathrm{assoc}}$ & 0.65 m \\
$\rho_{\mathrm{bright}},\rho_{\mathrm{dark}}$ & 0.60, 0.30 & $h_{\mathrm{clip}}$ & 0.15 m \\
$\tau_c$ & 1.5 s & $N_w,N_c$ & 30, 30 scans \\
$N_{\mathrm{cal}}$ & 4 observations & $\beta$ & 2 \\
$\tau_{\mathrm{mem}}$ & 1.5 s & $k_{\mathrm{w}}$ & 4 cells \\ $\mathcal N$ & 8 neighbors &  &  \\
\bottomrule
\end{tabular}

\end{table}

Both confirmation thresholds are set by the startup rule of \cref{sec:design:constants} with the windows of \cref{tab:notation}. For the simulator the 6\,km/h drive without an attack yields 25 eligible tracks with 254 track-frames, a largest track score of 7.0426 and a smallest track darkness of 0.2181, hence $T=14.085$ and $D_{\mathrm{FAST}}=0.1091$.
For the field sensor the opening scans of the 45\degree{} 0.18\,m$^2$ recording yield 2 eligible tracks with 5 track-frames, a largest score of 2.704 and a smallest darkness of 0.0975, hence $T=5.408$ and $D_{\mathrm{FAST}}=0.04875$. The simulator pair applied unchanged to all nine RW recordings reproduces 96.31\% recall and 98.30\% whole-frame precision (\cref{sec:eval:transfer}).
The window closes 8.2\,s before the first phantom-object return.
The labeled field phantom's track darkness (\cref{eq:kappat}) is 0.01, the sensor's lowest nonzero reflectivity code, against a threshold of 0.04875.

On the simulated drive without an attack no structure reaches a glint ratio above 2.35. $G_{\mathrm{MIN}}=3.0$ is the design constant above that maximum.
The real boards measure 4.4 to 13.4 over 248 board observations. Other structures also clear $G_{\mathrm{MIN}}$ on the real recordings, and precision rests on the opacity and darkness stages (\cref{tab:ablation}).
Over the simulated attack drives the phantom lies at most 12.78\,m behind the board plane, and $D_{\mathrm{IMG}}=19.0$\,m is 1.5 times that maximum rounded to half a meter.
On the RW recordings, varying one constant at a time, recall stays between 96.31 and 96.36\% for $D_{\mathrm{IMG}}$ from 12 to 25\,m, whole-frame precision stays between 98.28 and 98.64\%, and charged corridor returns rise from 1 to 6 above the frozen 19.0\,m (path-only precision 99.979\% at 25\,m). At 30\,m and beyond, the bound matters for precision, as the 45\degree{} 0.18\,m$^2$ recording falls to 71.51\%. In the same test, every other structural constant tested (the board minimum $n_{M,\min}$, the thickness bound $\sigma_{\max}$ and the height band $H_b$) leaves recall at 96.31\% and whole-frame precision at 98.30\%, except the raised lower height bound (0.3\,m), which reduces whole-frame precision to 97.43\%.
\Cref{tab:notation} lists the decision constants.

\section{Extended Results}
\label{app:results}

\subsection{Simulation and Labels}
\label{app:sim}
\Cref{tab:vsgeom} verifies the geometry of the virtual sensor at each tilt, and \cref{tab:simdata} lists the simulated drives by family.

\begin{table}[tb]
\caption{Geometric verification of the virtual-sensor fold, per tilt: maximum residual of the virtual sensor's pose against the mirror image of the real one, of the reflection law, and of the two-leg range, the beam-altitude residual, and whether every kept phantom cell crosses the aperture.}
\label{tab:vsgeom}
\centering
\tabfont
\begin{tabular}{@{}rrrrrl@{}}
\toprule
tilt & \multicolumn{3}{c}{residual max (m)} & beam-alt & cells in \\
\cmidrule(lr){2-4}
(deg) & pose & refl.\ law & range & med / max (deg) & aperture \\
\midrule
30 & $2.3\times10^{-5}$ & $4.3\times10^{-14}$ & $1.5\times10^{-5}$ & 0.108 / 0.180 & all \\
45 & $2.3\times10^{-5}$ & $7.1\times10^{-14}$ & $2.1\times10^{-5}$ & 0.077 / 0.907 & all \\
60 & $2.2\times10^{-5}$ & $5.7\times10^{-14}$ & $2.2\times10^{-5}$ & 0.052 / 1.043 & all \\
\bottomrule
\end{tabular}

\end{table}

\begin{table}[!tb]
\caption{Simulation drive inventory. Columns: drive family, sets, what varies, total drives, scans (attack plus clean twin), attack scans and clearly appearing scans (CA).}
\label{tab:simdata}
\centering
\tabfont
\begin{tabular}{@{}l r l r r r r@{}}
\toprule
drive family & sets & varied & drv. & scans & atk. & CA \\
\midrule
turn scenes & 6 & 3 tilts, 2 sources & 18 & 6606 & 3303 & 296 \\
speed repeats & 3 & 12 km/h & 8 & 2000 & 1000 & 88 \\
straight route & 9 & 20, 25, 30, 35 km/h & 45 & 6676 & 3338 & 180 \\
channel sweep & 8 & 16 to 128 channels & 32 & 11934 & 5967 & 288 \\
area sweep & 10 & 0.36 to 3.50 m$^2$ & 30 & 11120 & 5560 & 679 \\
\midrule
\textbf{total} & 36 &  & 133 & 38336 & 19168 & 1531 \\
\bottomrule
\end{tabular}

\end{table}

The hand labels of the RW recordings also confirm the aperture test of \cref{sec:physics:aperture}. Every labeled phantom return (object and background) on the 30\degree{} 0.60\,m$^2$ recording lies in the labeled board's footprint (1,343 of 1,343 over seven labeled frames). The aperture is necessary but not sufficient, as the \textit{Aperture only} column of \cref{fig:ablation} shows.

\subsection{Real-World Recordings}
\label{app:rw}

\Cref{tab:rwapp} gives 95\% block-bootstrap intervals ($B_{\mathrm{boot}}{=}10{,}000$, seed fixed) for recall and whole-frame precision. The block length is chosen per recording by a white-noise autocorrelation rule, capped at $\lfloor n/3\rfloor$.

\begin{table}[!tb]
\caption{Point-level results inside each recording's scored window. Pts: phantom-object points. Delay: scans from the first phantom scan to the first removal. Path: path-only precision to two decimals, pooled 99.996\%.}
\label{tab:rwapp}
\centering
\tabfont
\setlength{\tabcolsep}{2.4pt}
\begin{tabular}{@{}l r r r@{\,}l r@{\,}l r r@{}}
\toprule
 & \multicolumn{2}{c}{in window} & \multicolumn{2}{c}{recall (\%)} & \multicolumn{2}{c}{whole prec.\ (\%)} & path & delay \\
\cmidrule(lr){2-3}\cmidrule(lr){4-5}\cmidrule(lr){6-7}
recording & scans & pts & & [95\% CI] & & [95\% CI] & (\%) & \\
\midrule
30\degree{} 0.18 & 18 & 1893 & 95.51 & [91.8,\,97.9] & 99.88 & [99.7,\,100.0] & 100.00 & 0 \\
30\degree{} 0.36 & 25 & 5412 & 96.32 & [95.3,\,97.7] & 99.95 & [99.8,\,100.0] & 99.98 & 0 \\
30\degree{} 0.60 & 30 & 7485 & 97.22 & [96.5,\,97.9] & 98.86 & [97.4,\,99.7] & 100.00 & 1 \\
45\degree{} 0.18 & 17 & 484 & 99.17 & [96.5,\,100.0] & 100.00 & [100.0,\,100.0] & 100.00 & 1 \\
45\degree{} 0.36 & 25 & 1915 & 93.47 & [90.6,\,96.4] & 94.24 & [83.3,\,100.0] & 100.00 & 1 \\
45\degree{} 0.60 & 23 & 2383 & 95.97 & [94.0,\,97.9] & 96.01 & [89.8,\,100.0] & 100.00 & 0 \\
60\degree{} 0.18 & 25 & 191 & 100.00 & [100.0,\,100.0] & 100.00 & [100.0,\,100.0] & 100.00 & 0 \\
60\degree{} 0.36 & 27 & 831 & 97.95 & [93.1,\,99.9] & 99.54 & [98.7,\,100.0] & 100.00 & 1 \\
60\degree{} 0.60 & 34 & 1243 & 94.29 & [89.4,\,98.1] & 95.73 & [90.5,\,99.2] & 100.00 & 1 \\
\midrule
\textbf{pooled} & 224 & 21837 & 96.31 & & 98.30 & & 100.00 & \\
\multicolumn{3}{@{}l}{mean $\pm$ SD} & \multicolumn{2}{l}{96.7$\pm$2.2} & \multicolumn{2}{l}{98.2$\pm$2.3} & \multicolumn{2}{l@{}}{100.00$\pm$0.01} \\
\bottomrule
\end{tabular}

\end{table}

\begin{figure}[!tb]
\centering
\includegraphics[width=\linewidth]{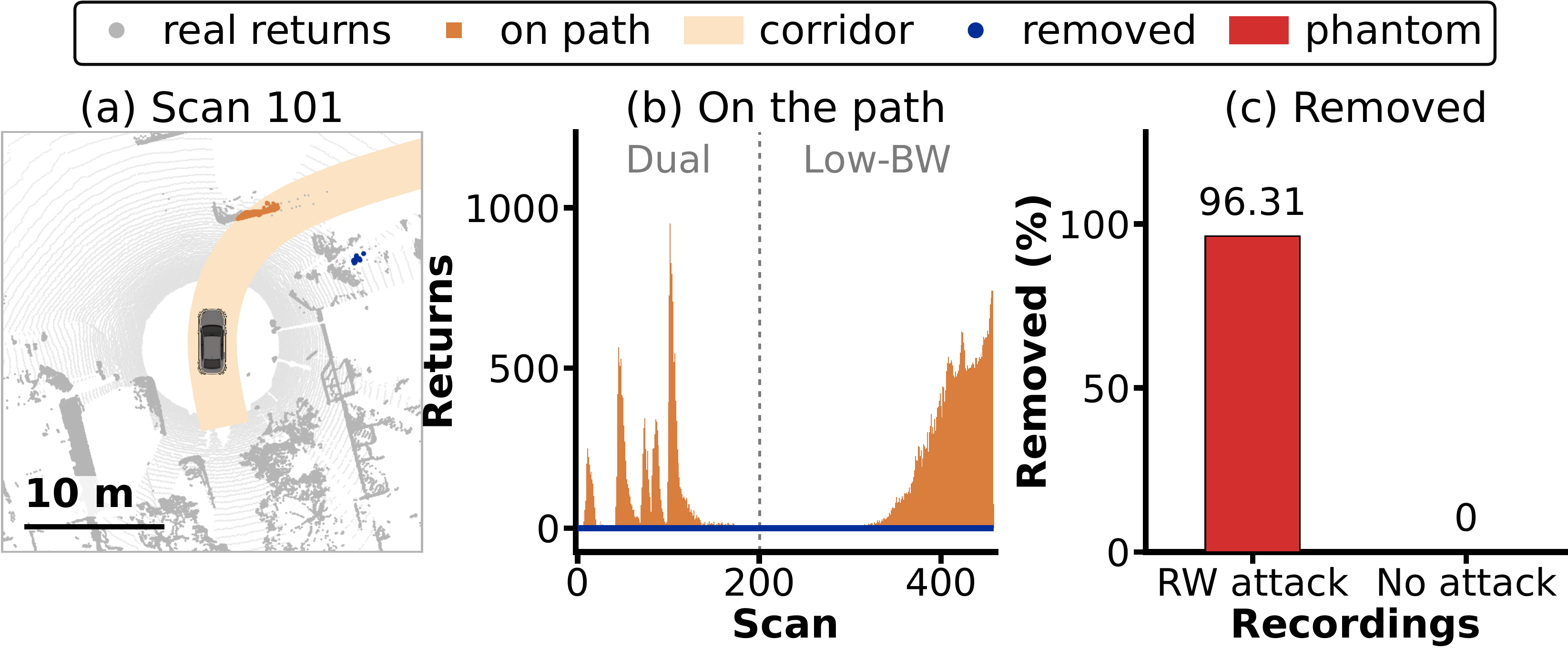}
\caption{Attack-free public driving on two OS1-128 city-driving recordings~\cite{ouster2022samples} (459 scans, no attack), frozen field constants. (a)~One scan, bird's-eye. (b)~Object-height returns on the driven path per scan. (c)~Share removed: phantom-object returns of the nine attacked RW recordings (left) against on-path object-height returns of the two public drives (right).}
\label{fig:public}
\end{figure}

\system{} is also tested on attack-free data from another sensor unit and a city street. On two public OS1-128 city-driving recordings (459 scans, Dual-Returns and Low-Bandwidth profiles)~\cite{ouster2022samples}, with the frozen field constants unchanged, \system{} removes no return on the driven corridor under every threshold pair tested (the field pair, the simulator pair and each drive's own startup pair).
Off the path, at most 0.0134\% of the returns are removed. The three removal events are dark real objects. Of 64,006 object-height returns on the driven path (0.5 to 3\,m above the road, outside the ego body), where the audited scans show a lead vehicle in the lane, none is removed (\cref{fig:public}).

\subsection{Sensitivity to the Constants}
\label{app:sens}

\begin{table}[!tb]
\caption{Sensitivity of the RW result to the confirmation constants. Rows 1 to 9: each recording's opening window, all nine scored. $^{*}$\,Frozen pair. Rows 10 and 11: worst value over the stated range. Path cells: charged corridor returns. Path precision: two decimals, pooled 99.996\%.}
\label{tab:robust}
\centering
\tabfont
\begin{tabular}{@{}lrrrrrr@{}}
\toprule
calibration & $T$ & $D_{\mathrm{FAST}}$ & recall & prec. & prec. & path \\
window & & & (\%) & whole (\%) & path (\%) & cells \\
\midrule
30\degree{} 0.18 & 25.511 & 0.015 & 95.23 & 98.53 & 100.00 & 1 \\
30\degree{} 0.36 & 27.661 & 0.015 & 95.23 & 98.53 & 100.00 & 1 \\
30\degree{} 0.60 & 28.981 & 0.015 & 95.23 & 98.53 & 100.00 & 1 \\
45\degree{} 0.18$^{*}$ & 5.408 & 0.04875 & 96.31 & 98.30 & 100.00 & 1 \\
45\degree{} 0.36 & 27.146 & 0.06 & 96.36 & 98.30 & 100.00 & 1 \\
45\degree{} 0.60 & 17.663 & 0.06 & 96.36 & 98.30 & 100.00 & 1 \\
60\degree{} 0.18 & 27.506 & 0.055 & 96.36 & 98.30 & 100.00 & 1 \\
60\degree{} 0.36 & 27.205 & 0.02 & 96.36 & 98.69 & 100.00 & 1 \\
60\degree{} 0.60 & 27.449 & 0.025 & 96.36 & 98.69 & 100.00 & 1 \\
\midrule
both scaled & \multicolumn{2}{c}{$\times$0.5 to 2} & 96.31 & 97.73 & 100.00 & 1 \\
$G_{\mathrm{MIN}}$ 1.5 to 6 & 5.408 & 0.04875 & 96.31 & 97.44 & 100.00 & 1 \\
simulator & 14.085 & 0.1091 & 96.31 & 98.30 & 100.00 & 1 \\
without 45\degree{} 0.18 & 5.408 & 0.04875 & 96.25 & 98.26 & 100.00 & 1 \\
\bottomrule
\end{tabular}

\end{table}

Each row of \cref{tab:robust} pools the nine recordings.

\begin{figure}[!t]
\centering
\includegraphics[width=\columnwidth]{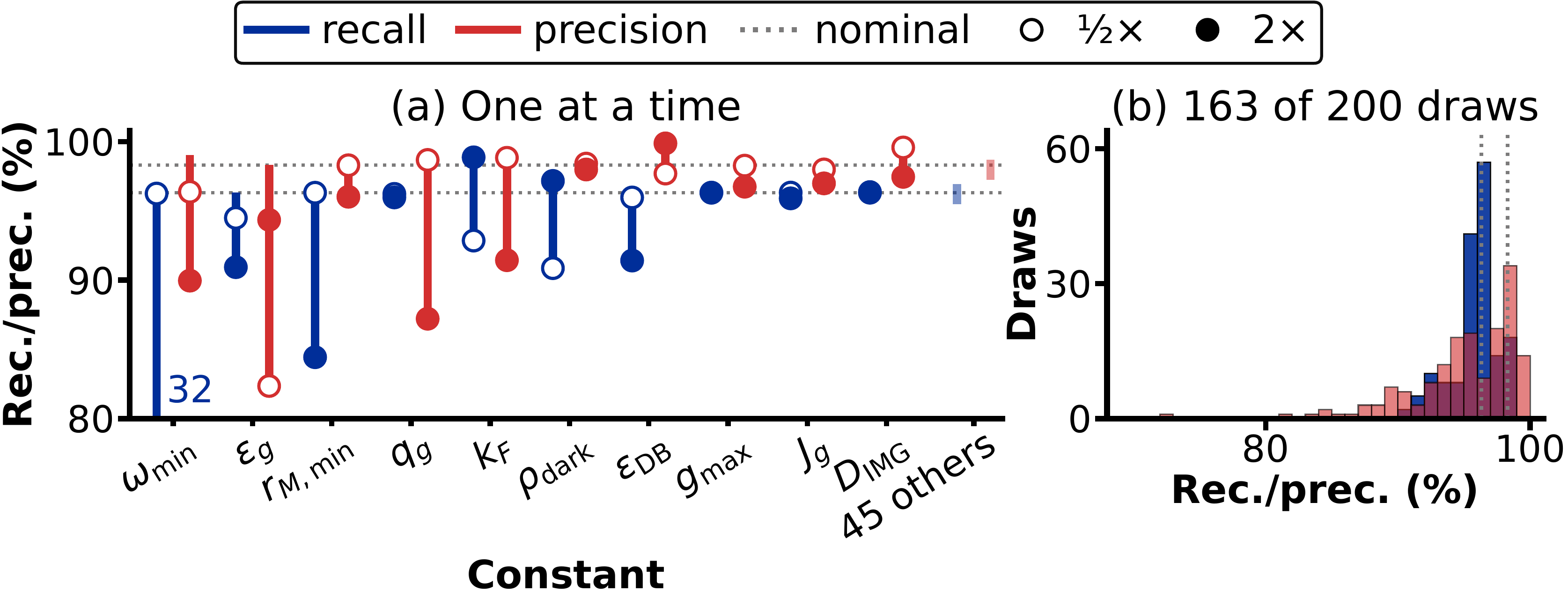}
\caption{Sensitivity of the RW result to each of the 55 constants, from half to twice the nominal value, dotted: nominal configuration. Recall (blue) and precision (red). (a)~Ten largest excursions and envelope of the 45 others. (b)~The 163 of 200 joint draws in which the filter starts, each constant between 0.75 and 1.25 times its nominal value. No value was selected.}
\label{fig:sweep}
\end{figure}

Each constant is moved to 0.5, 0.75, 1.5 and 2 times its nominal value, with integer constants rounded to the nearest admissible value and bounded constants stopped at their domain (\cref{fig:sweep}).
The 55 constants are the 37 rule and fixed design constants in \cref{tab:notation} (the startup pair, the sensor constants, the neighborhood $\mathcal N$, the sparse threshold and the micro alarm were not moved, and the two release parameters are separate) and 18 implementation constants of the ground fit, the clustering, the glint band, the track linking and the carry of \cref{sec:design}, five in \cref{fig:sweep}: the ground tolerance $\epsilon_g = 0.25$\,m, the ground fraction $q_g = 0.40$, the RANSAC budget $J_g = 50$, the clustering radius $\epsilon_{\mathrm{DB}} = 0.7$\,m and the link gap $g_{\max} = 3$ scans. The implementation settings of the clustering, ground fit, linking, tracker and darkness bins are fixed.
The nominal value $\omega_{\min}=0.5$ requires at least half of a see-through cluster's cells to lie inside the board's pixel footprint (\cref{eq:overlap}), and at $\omega_{\min}=0.75$ (1.5 times the nominal value, three quarters of the cells) recall is 96.26\% at 99.05\% precision. At $\omega_{\min}=1$ (leftmost in~(a)), the largest value the share can take, every cell must lie inside the footprint. The footprint is the board's own returned cells dilated by $k_F$ and the clusters are formed over the whole non-ground scan, so a cluster with a single cell outside the footprint fails the test at $\omega_{\min}=1$, and recall drops to 31.68\% at 89.96\% precision.
Three startup constants ($N_w$, $N_c$, $N_{\mathrm{cal}}$) have single moves at which the startup rule returns not calibrated on the field opening window and the filter does not start (5 moves).

\subsection{Baselines and Cost}
\label{app:learned}

\begin{table}[!tb]
\caption{Per-seed recall and precision of the five learned baselines, for the validation-selected arm and the two fixed input variants. Seeds 1909, 1919, 1929, each pooled over the nine test recordings.}
\label{tab:learnedseed}
\centering
\tabfont
\begin{tabular}{@{}lccccccc@{}}
\toprule
Model (input & \multicolumn{3}{c}{Recall (\%), seed} & Recall & \multicolumn{3}{c}{Precision (\%), seed} \\
variant) & 1909 & 1919 & 1929 & mean$\pm$SD & 1909 & 1919 & 1929 \\
\midrule
LightGBM (selected) & 56.81 & 54.46 & 56.22 & 55.83$\pm$1.22 & 32.84 & 35.56 & 35.84 \\
\quad raw & 56.40 & 55.69 & 56.09 & 56.06$\pm$0.36 & 33.25 & 35.36 & 35.47 \\
\quad log-std & 56.40 & 54.33 & 56.23 & 55.65$\pm$1.15 & 32.96 & 36.30 & 35.84 \\
PointNet (selected) & 51.38 & 47.72 & 44.24 & 47.78$\pm$3.57 & 42.92 & 44.37 & 44.08 \\
\quad raw & 49.04 & 47.78 & 45.07 & 47.29$\pm$2.03 & 46.19 & 39.93 & 42.67 \\
\quad log-std & 47.81 & 46.98 & 40.86 & 45.22$\pm$3.80 & 46.11 & 49.70 & 45.62 \\
WeatherNet (selected) & 73.58 & 73.83 & 63.17 & 70.19$\pm$6.08 & 63.27 & 65.06 & 59.24 \\
\quad raw & 74.08 & 73.98 & 62.93 & 70.33$\pm$6.40 & 62.92 & 65.00 & 59.27 \\
\quad log-std & 67.39 & 62.61 & 57.30 & 62.43$\pm$5.05 & 59.64 & 59.98 & 54.00 \\
SalsaNext (selected) & 97.96 & 72.62 & 95.74 & 88.77$\pm$14.03 & 99.40 & 99.71 & 98.52 \\
\quad raw & 95.25 & 79.37 & 94.88 & 89.84$\pm$9.06 & 99.01 & 97.09 & 97.59 \\
\quad log-std & 98.55 & 84.22 & 77.87 & 86.88$\pm$10.60 & 98.39 & 99.69 & 99.19 \\
LMNet (selected) & 98.32 & 98.15 & 98.66 & 98.38$\pm$0.26 & 94.25 & 97.72 & 99.68 \\
\quad raw & 98.68 & 97.88 & 99.67 & 98.74$\pm$0.90 & 99.18 & 93.92 & 98.64 \\
\quad log-std & 89.84 & 89.00 & 89.69 & 89.51$\pm$0.45 & 93.88 & 99.00 & 98.65 \\
\bottomrule
\end{tabular}

\end{table}

\Cref{tab:learnedseed} disaggregates the learned baselines of \cref{tab:learned} by training seed.
Each model is trained in nine leave-one-recording-out folds with three seeds per fold (270 trainings over the five models and two variants).
Each network keeps the final epoch of its published schedule.
The decision threshold is the middle of the range of thresholds that reach the maximum point-wise F1 on the validation recording.
Each model reads reflectivity either raw or log-standardized, and in each fold the log-standardized variant is used only when its mean validation F1 over the three seeds is higher.
The table prints the selected variant and both fixed variants.
LightGBM follows the training recipe and settings of Sevgen and Abdikan~\cite{sevgen2023lightgbm} on seven per-point features: range, height above the estimated ground, density of range-matched neighbors in a $5\times 5$ window, log reflectivity minus its local mean, local reflectivity spread, beam index and height over range.
A model never reads the labels of its test recording.
LMNet reads eight residual images of past scans with poses from KISS-ICP.
LightGBM fits 0.10 million tree nodes, while PointNet, WeatherNet, SalsaNext and LMNet learn 1.17, 1.53, 6.71 and 6.71 million parameters.
Measured like the latency (8 threads, one L40S, back-to-back) and net of idle power, the energy per scan is 4.5\,J for \system{} on the CPU, against 10.3 to 48.2\,J for the five models on the CPU and 4.3 to 16.7\,J for the three networks measured on the GPU.
PointNet's retired instructions are not counted.
For the Unseen tilt column of \cref{tab:learned}, the five models follow the rules above with three seeds but are trained without one tilt, whose three recordings are scored.
Of the six recordings of the two other tilts, the 0.36\,m$^2$ recording of one sets the threshold and the other five fit the model, so each fold learns from five recordings instead of seven.
\system{} is compared on the 30\degree{} and 60\degree{} recordings only, because its $T$ and $D_{\mathrm{FAST}}$ come from the opening scans of a 45\degree{} recording (\cref{sec:setup}).

The injection-defense cues of \cref{sec:eval:baselines} are computed from their published definitions.
They are evaluated on 191 labeled RW phantom objects (at least 10 labeled points each) and on the simulated drives (569 phantom and 5,500 genuine-vehicle reference objects), each in a tight bounding box and a full-car box ($4.5\times 1.8\times 1.5$\,m, extended away from the sensor).
CARLO's thresholds are read off the published figures of Sun et~al.~\cite{sun2020towards}. Shadow-Catcher's unpublished weight $\alpha$ is swept from 0.1 to 10 (start line at the farthest box corner). ADoPT's temporal cluster consistency uses an ideal object flow and a per-object decision in place of the scene flow ADoPT fits online for each frame pair and its frame-level metric.
No genuine-vehicle label exists on the RW recordings, so the RW phantom is read against the simulated vehicles.
Per test, CARLO's cue flags 0 of the 191 RW phantom objects on a tight bounding box (2 of 191 on a full-car box) and 0 of 5,500 simulated genuine vehicles on the tight box. ADoPT's cue flags 0 of 191. The shadow region of Shadow-Catcher flags genuine vehicles at a rate of the same order as the phantom at every setting of its unpublished weight, without reaching separation: at $\alpha{=}1$, 100\% of the range-matched simulated vehicles against 68.1\% of the phantom objects on a tight box, because the published region behind a genuine vehicle is not empty in these scans.

\Cref{tab:roc} gives the per-recording AUC and true-positive rate at 5\% false-positive rate.

\begin{table}[!tb]
\caption{Point-level detection on the RW recordings, phantom-object points against up to 4000 other returns per scan, inside the scored windows. The naive intensity threshold of the reference attack's authors~\cite{yahia2025seeing}, which flags bright returns as artifacts, against a per-point darkness score.}
\label{tab:roc}
\centering
\tabfont
\begin{tabular}{@{}lrrrrr@{}}
\toprule
recording & phantom & \multicolumn{2}{c}{ROC AUC} & \multicolumn{2}{c}{TPR at 5\% FPR} \\
\cmidrule(lr){3-4}\cmidrule(lr){5-6}
 & points & naive & dark & naive & dark \\
\midrule
30\degree{} 0.18 & 1893 & 0.125 & 0.859 & 0.009 & 0.500 \\
30\degree{} 0.36 & 5412 & 0.115 & 0.880 & 0.005 & 0.598 \\
30\degree{} 0.60 & 7485 & 0.114 & 0.881 & 0.006 & 0.611 \\
45\degree{} 0.18 & 484 & 0.161 & 0.861 & 0.006 & 0.465 \\
45\degree{} 0.36 & 1915 & 0.140 & 0.881 & 0.010 & 0.594 \\
45\degree{} 0.60 & 2383 & 0.140 & 0.881 & 0.009 & 0.550 \\
60\degree{} 0.18 & 191 & 0.172 & 0.850 & 0.010 & 0.372 \\
60\degree{} 0.36 & 831 & 0.181 & 0.842 & 0.065 & 0.546 \\
60\degree{} 0.60 & 1243 & 0.198 & 0.827 & 0.074 & 0.524 \\
\midrule
pooled & 21837 & 0.132 & 0.875 & 0.013 & 0.594 \\
mean $\pm$ SD & & 0.150$\pm$0.030 & 0.862$\pm$0.020 & 0.022$\pm$0.027 & 0.529$\pm$0.076 \\
\bottomrule
\end{tabular}

\end{table}

\Cref{tab:runtime} gives the quiet-machine thread matrix (131,072 cells per scan, 104,777 returns on average). On one RW recording at eight threads, a mean of 2.3 board candidates (geometric test, before the glint gate) and 10.7 candidate records per scan keep the per-board terms small.
\system{} as shipped with decoding, measured at 10\,Hz over three passes, falls at most 0.08\,s behind after the start-up (p99 0.05\,s), against 0.08\,s projected for the same runs.

\begin{table}[!tb]
\caption{Cost per scan of \system{} by thread count, pooled over the nine RW recordings, on a quiet Intel(R) Xeon(R) Gold 5420+. The 8-thread row pools three passes. The 8\,+\,release row adds the release step.}
\label{tab:runtime}
\centering
\tabfont
\begin{tabular}{@{}rcccccc@{}}
\toprule
Threads & Mean & Median & p95 & p99 & $<$100\,ms & Full path \\
 & ms & ms & ms & ms & \% & ms \\
\midrule
1 & 93 $\pm$ 3 & 93 & 114.8 & 129.7 & 72.3 & 104.6 \\
2 & 74 $\pm$ 3 & 72 & 94.8 & 110.3 & 97.4 & 85.2 \\
4 & 62 $\pm$ 2 & 60 & 83.0 & 98.4 & 99.2 & 73.1 \\
8 & 56 $\pm$ 2 & 53 & 77.1 & 92.4 & 99.5 & 67.3 \\
16 & 64 $\pm$ 3 & 61 & 91.8 & 109.7 & 97.8 & 77.4 \\
\midrule
8 + release & 59 $\pm$ 3 & 56 & 84.5 & 103.2 & 98.9 & 70.4 \\
\bottomrule
\end{tabular}

\end{table}

\subsection{Downstream Perception}
\label{app:downstream}

On the six 30\,km/h straight-route drives in which Autoware's shipped tracker publishes a phantom CAR without the defense (\cref{fig:downstream}b), the existence probability of the phantom track reaches 0.769 to 0.992.
On 4 of these drives the first published message follows the phantom's last scan.
Every phantom track that the shipped tracker publishes on the straight route lies at least 5.9\,m beside the driven path.

\subsection{Component Ablation}
\label{app:ablation}

\begin{figure}[!tb]
\centering
\includegraphics[width=\columnwidth]{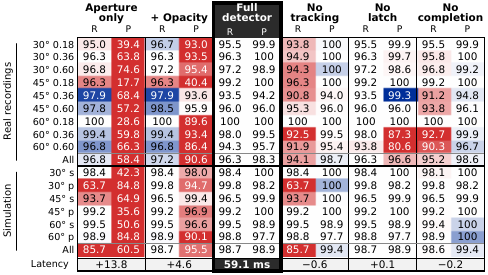}
\caption{Component ablation. Rows: the nine RW recordings (tilt, board area in m$^2$) and six simulated scenes (s~self-image, p~parked source), pooled in All. Columns: recall~(R) and precision~(P) per variant. Blue marks a value above the full detector, red below, white equal, saturating at 5 points. Latency in ms: the full detector's per-scan cost and each variant's~difference.}
\label{fig:ablation}
\label{tab:ablation}
\end{figure}

Each column of \cref{fig:ablation} changes the full detector as follows: \textit{Aperture only} alarms every candidate at its own time with band, support and opacity bypassed. \textit{+ Opacity} restores the opacity gate to it. \textit{No tracking} disables the tracker and bracketed carry. \textit{No latch} disables the track latch (digest latch stays). \textit{No completion} disables core completion. The aperture-only variant costs 13.8\,ms more per scan, consistent with the opacity and darkness stages discarding candidates early, and with the opacity gate restored the variant still costs 4.6\,ms more. Tracking, latch and completion each move the cost by less than 1\,ms.

\end{document}